\documentclass[review]{cas-sc} 
\usepackage[authoryear]{natbib}

\usepackage{graphicx}
\usepackage{xspace}
\usepackage{natbib}
\usepackage{lscape}
\usepackage{rotating}
\usepackage{array}
\usepackage{lscape}

\bibpunct{(}{)}{;}{a}{}{,}
\usepackage{hyperref}
\hypersetup{colorlinks=true, linktocpage=true, pdfstartpage=3,
            pdfstartview=FitV, breaklinks=true, pdfpagemode=UseNone,
            pageanchor=true, pdfpagemode=UseOutlines, plainpages=false,
            bookmarksnumbered, bookmarksopen=true, bookmarksopenlevel=1,
            hypertexnames=true, pdfhighlight=/O, urlcolor=RoyalBlue,
            linkcolor=RoyalBlue, 
            }

\newcommand{\source}[1]{\textsuperscript{\textcolor{blue}{[citation needed]}}\xspace}

\usepackage[switch]{lineno}

\begin{document}
\let\WriteBookmarks\relax
\def\floatpagepagefraction{1}
\def\textpagefraction{.001}
\shorttitle{Asteroid-Meteorite Connections}
\shortauthors{DeMeo, F. E. et~al.}

\title [mode = title]{Connecting Asteroids and Meteorites with visible and near-infrared spectroscopy}                      

\author[mit]{Francesca E. DeMeo}[orcid=0000-0002-8397-4219] \cormark[1]
\author[mit,low]{Brian J. Burt}[orcid=0000-0002-6423-0716]
\author[mit,eso]{Micha\"el Marsset}[orcid=0000-0001-8617-2425]
\author[wei]{David Polishook}[orcid=0000-0002-6977-3146]
\author[mhc]{Thomas H. Burbine}[orcid=0000-0001-8889-8692]
\author[nic]{Beno\^it Carry}  [orcid=0000-0001-5242-3089]
\author[mit]{Richard P. Binzel}[orcid=0000-0002-9995-7341]
\author[mar]{Pierre Vernazza}  [orcid=0000-0002-2564-6743]
\author[lpl]{Vishnu Reddy} [orcid=0000-0002-7743-3491] 
\author[mit]{Michelle Tang}
\author[nau]{Cristina A. Thomas}[orcid=0000-0003-3091-5757]
\author[apl]{Andrew S. Rivkin}[orcid=0000-0002-9939-9976]
\author[low]{Nicholas A. Moskovitz}[orcid=0000-0001-6765-6336]
\author[mit]{Stephen M. Slivan}[orcid=0000-0003-3291-8708]
\author[ifa]{Schelte J. Bus}[orcid=0000-0003-4191-6536]

\address[mit]{Department of Earth, Atmospheric, and Planetary Sciences, Massachusetts Institute of Technology, 77 Massachusetts Avenue, Cambridge, MA 02139 USA}
\address[low]{Lowell Observatory, 1400 West Mars Hill Road, Flagstaff, AZ 86001, USA}
\address[eso]{European Southern Observatory (ESO), Alonso de Cordova 3107, 1900 Casilla Vitacura, Santiago, Chile}
\address[wei]{Faculty of Physics, Weizmann Institute of Science, Rehovot 0076100, Israel}
\address[mhc]{Department of Astronomy, Mount Holyoke College, South Hadley, MA 01075, USA}
\address[nic]{Observatoire de la C{\^o}te d'Azur, Boulevard de l'Observatoire, 06304 Nice Cedex 4, France}
\address[mar]{Aix Marseille Universit{\'e}, CNRS, CNES, Laboratoire d’Astrophysique de Marseille, Marseille, France}
\address[lpl]{Lunar and Planetary Laboratory, University of Arizona, 1629 E University Blvd, Tucson, AZ 85721, USA}
\address[nau]{Northern Arizona University, Department of Astronomy and Planetary Science PO Box 6010, Flagstaff, AZ 86011 USA}
\address[apl]{The Johns Hopkins University Applied Physics Laboratory, Laurel, MD, USA}
\address[ifa]{Institute for Astronomy, University of Hawaii, 2860 Woodlawn Drive, Honolulu, HI  96822-1839, USA}

\cortext[cor1]{Corresponding author, fdemeo@mit.edu}

\begin{abstract}
In this work we identify spectral similarities between asteroids and meteorites. Using spectral features such as absorption bands and spectral curvature, we identify spectral matches between 500 asteroid spectra and over 1,000 samples of RELAB meteorite spectra over visible plus near-infrared wavelengths (0.45--2.5 $\mu$m).  We reproduce and confirm many major and previously known meteorite-asteroid connections and find possible new, more rare or less-established connections. Well-established connections include: ordinary chondrites with S-complex asteroids; pristine CM carbonaceous chondrites with Ch-type asteroids and heated CMs with C-type asteroids ; HED meteorites with V-types; enstatite chondrites with Xc-type asteroids; CV meteorites with K-type asteroids; Brachinites, Pallasites and R chondrites with olivine-dominated A-type asteroids.

In addition to the link between ordinary chondrite meteorites with S-complex asteroids, we find a trend from Q, Sq, S, Sr to Sv correlates with LL to H, with Q-types matching predominately to L and LL ordinary chondrites, and Sr and Sv matching predominantly with L and H ordinary chondrites.  We find ordinary chondrite samples that match to the X-complex. These are measurements of slabs and many are labeled as dark or black (shocked) ordinary chondrites. We find carbonaceous chondrite samples having spectral slopes large enough to match D-type asteroid spectra. 

We find in many cases the asteroid type to meteorite type links are not unique, for classes with and without distinct spectral features. While there are examples of dominant matches between an asteroid class and meteorite class that are well established, there are less common but still spectrally compatible matches between many asteroid types and meteorite types. This result emphasizes the diversity of asteroid and meteorite compositions and highlights the degeneracy of classification by spectral features alone requiring additional measurements to firmly establish asteroid-meteorite links. Recent and upcoming spacecraft missions will shed light on the compositions of many of the asteroid classes, particularly those without diagnostic features, (C-, B-, X-, and D-types), with measurements of C-type Ceres, C-type Ryugu, B-type Bennu, M-type Psyche, and C-, P-, and D-types as part of the Lucy mission.

\end{abstract}


\begin{highlights}
\item We search for spectral links among asteroid and meteorite classes by comparing 500 asteroid spectra and over 1,000 samples of meteorites over visible plus near-infrared (0.45--2.5 $\mu$m) wavelengths.
\item We confirm established links, and also find less common but spectrally compatible matches between many asteroid types and meteorite types.
\item In some cases, multiple asteroid-meteorite links emphasize the diversity of compositions, and highlight the degeneracy of classification by spectral features alone.

\end{highlights}


\maketitle


\section{Introduction}
More than a million asteroids larger than 1 km derived from hundreds of parent bodies preserve records of the early Solar System conditions. Those parent bodies are expected to include existing large ($>\sim$100 km) bodies \citep{2009Icar..204..558M}, and ones that were subsequently disrupted. Our meteorite collections have evidence for approximately 100-150 parent bodies \citep{Greenwood2020}. Each asteroid is unique in composition, shape, size, and other physical properties, although they are generally grouped into a couple dozen broad spectral classes \citep{Tholen1984, Bus2002b, DeMeo2009taxo}.  An impressive body of knowledge of asteroid compositions has been gathered through remote sensing and spacecraft missions. Even so, the level of detail that can be obtained through the study of meteorites in laboratories on Earth is incomparable. Thus, linking meteorites to asteroids is critical to unlocking the wealth of compositional information contained in meteorites.

Forging these asteroid-meteorite connections is not trivial because their spectra are rarely straightforward one-to-one matches. The asteroid measurements are averaged over a surface meters to hundreds of kilometers across while meteorite measurements are taken on a small, localized sample. Spectra also contain more than compositional information, they are dependent on factors such as the grain size of the particles on the asteroid surface or meteorite sample, phase angle of the observation, and temperature \citep{1973JGR....78.8507J,1981JGR....86.4571H,Roush1984,1992LPSC...22..313H,1981JGR....86.4571H,Reddy2015,2016AJ....152...54V,Marsset2020}. Asteroid surfaces are also affected by the space environment, a process called space weathering \citep{1965NYASA.123..711H, 2001JGR...10610039H, 2006Natur.443...56H,2011Sci...333.1121N,Brunetto2015}, and their measurements are affected by Earth observing conditions of the night such as the airmass \citep{Marsset2020}. Additionally, as demonstrated by a study of HED meteorites and by NASA's Dawn mission to asteroid Vesta, asteroid surfaces are rarely pristine. Hemispherical scale contamination by exogenic material makes asteroid-meteorite links challenging \citep{Reddy2012c}. As one example, whether a measured meteorite sample is a particulate (grains rather than slabs or chips) or raw or polished chips or slabs can affect slope, absorption band depth, shape and overall reflectance levels \citep[e.g.,][]{Cloutis2011,KramerRuggiu2021}. Despite these challenges, much work has been done to forge these connections at both levels of individual asteroids and meteorites \citep[e.g.,][]{MotheDiniz2005, Reddy2012a, Reddy2012b, DeLeon2004, Popescu2012} and more broadly for asteroid classes to meteorite classes \citep[e.g.,][]{Gaffey1993, Thomas2010, Vernazza2008, Rivkin2015, Vernazza2015, Vernazza2017, Burbine2002AIII, Burbine2014, Greenwood2020, Saito2020,KramerRuggiu2021,Eschrig2021} 

Here we take the largest available visible+near-infrared (0.45--2.5 $\mu$m) spectral dataset of asteroids, 500 spectra from the MIT-Hawaii Near-Earth Object Survey \citep[MITHNEOS,][]{Binzel2004, Binzel2019, Marsset2022} and the Small Main Belt Asteroid Spectroscopic Survey \citep[SMASS,][]{ Bus2002a, DeMeo2009taxo} as well as from other works (see details in Sec~\ref{Sec:Data}) and compare it to the largest publicly available dataset of meteorites - the Reflectance Experiment Laboratory database \citep[RELAB] {Pieters2004,2016LPI....47.2058M}. We seek to compare these datasets to validate existing asteroid-meteorite connections and explore new connections.

\section{Data and Taxonomic Classification}
\subsection{Asteroid and Meteorite Spectral Data} \label{Sec:Data}
For asteroids, we include 500 visible and near-infrared spectra most of which are publicly available from the SMASS and MITHNEOS programs \citep{Bus2002a, Binzel2004, DeMeo2009taxo, Binzel2019, Marsset2022} as well as other published literature. The full list of data references is provided in Table~\ref{Table:DataRefs}, and a full list of asteroids with their observation dates and visible and near-infrared data references are provided in the Supplementary Data. Most visible-wavelength observations were made between August 1993 and March 1999 using the 2.4-m (f/7.5) Hiltner or 1.3-m (f/7.6) McGraw–Hill reflecting telescopes at the MDM Observatory. The near-infrared data were obtained using SpeX on the NASA IRTF \citep{Rayner2003}. MITHNEOS data collected through 2020 were included in the analysis. A number of asteroids had near-infrared spectral measurements from multiple nights, in which case we use the spectrum with the highest signal-to-noise ratio. We restrict our sample to asteroids having spectra with signal-to-noise ratios greater than 30 at 2-$\mu$m that have both visible and near-infrared data available. Spectral plots of the 500 asteroid spectra are provided in the Supplementary Data. These data were given taxonomic assignments in previous work \citep{Bus2002a, Binzel2004, DeMeo2009taxo, Binzel2019, Marsset2022}, and we use the existing classifications for these spectra in this work. We include the following types from the Bus-DeMeo \citep{DeMeo2009taxo} classification system: S-complex: Q, S, Sq, Sr, Sv, C-complex: B, C, Cb, Ch (Ch+Cgh), X-complex: X (X+T), Xc, Xe, Xk, Xn, and additional classes D, L, K, A (A+Sa), V. We combine A and Sa results since their spectra are distinguished only by slope. We combine Ch and Cgh results because they both display a 0.7-$\mu$m feature that is the dominant spectral characteristic of those classes. We combine the X and T classes because they do not have significant differences over the full visible and near-infrared wavelengths. We do not include the Cg, R, or O classes for which we had a single asteroid spectrum for each class. 

\begin{table}[width=.9\linewidth,cols=3,pos=h]
\caption{Asteroid Spectroscopic Data References Used in this Work}
\label{Table:DataRefs}
\begin{tabular*}{\tblwidth}{L }
\toprule
Asteroid Spectroscopic Data References	\\
\midrule											
\citet{1985Icar...61..355Z,1990AJ.....99.1985L,1995Icar..115....1X,1998Icar..133...69H} \\
\citet{2001Icar..151..139B,2001PhDT.......121W,Binzel2001,Burbine2002} \\
\citet{Bus2002a,2004Icar..169..373L,2004MPS...39..351B,Binzel2004} \\
\citet{2006AA...451..331M,2006PhDTVernazza,2007Icar..186..111D,2007Icar..192..469R} \\
\citet{2007LPI....38.1104G,2008Icar..193...20S,2008MSAIS..12...20L,DeMeo2009taxo} \\
\citet{2009ATel.2323....1H,2009Icar..200..480B,2010AA...517A..23D,2010ATel.2488....1H} \\
\citet{2010ATel.2571....1H,2010ATel.2822....1H,2010ATel.2859....1H,2010Icar..208..773M} \\
\citet{2011AA...535A..15P,2011ATel.3678....1H,2011Icar..216..462C,2011PDSS..145.....H} \\
\citet{2012ATel.4251....1H,2013ATel.5132....1H,2013Icar..225..131S,2013LPI....44.1813L} \\
\citet{2014AA...569A..59I,2014Icar..228..217T,2014PSS...92...57R,2014PASJ...66...51K} \\
\citet{2016AJ....151...11P,2016Icar..268..340C,Binzel2019,Marsset2022} \\

\bottomrule
\end{tabular*}
\end{table}

For meteorites, we used the RELAB online database at Brown University \citep{Pieters2004,2016LPI....47.2058M} downloaded on Dec 16, 2019. We restrict this sample to spectral measurements that have data from 0.45 through 2.45 $\mu$m to overlap with the asteroid spectra. In this wavelength range, RELAB includes more than 20,000 spectra of 8,000 samples defined as the "Spectrum ID" and first 10 digits of the "Sample ID" in the RELAB catalog. Of these, more than 3,000 spectra of over 1,000 samples were meteorites (see Table~\ref{Table:NType1}). 

\begin{table}[width=.9\linewidth,cols=3,pos=h]
\caption{Number of Samples and Spectra of Meteorite Types and Subtypes in RELAB Used in this Study}
\label{Table:NType1}
\begin{tabular*}{\tblwidth}{LLLLLL }
\toprule
Meteorite	&	Subtype	&	N Sample	&	N Sample	&	N Spectra	&	N Spectra	\\
Type	&		&	All	&	Particulates	&	All	&	Particulates	\\
\midrule											
Achondrite (AC)	&		&	347	&	297	&	1073	&	941	\\
	&	Acapulcoite-Lodranite	&	30	&	18	&	94	&	67	\\
	&	Almahata Sitta	&	19	&	14	&	95	&	81	\\
	&	Aubrite	&	16	&	14	&	64	&	58	\\
	&	Brachinite	&	12	&	11	&	49	&	44	\\
	&	Diogenite	&	35	&	31	&	135	&	124	\\
	&	Eucrite	&	110	&	96	&	294	&	261	\\
	&	Howardite	&	83	&	77	&	242	&	231	\\
	&	Other AC	&	17	&	13	&	42	&	24	\\
	&	Ureilite	&	23	&	21	&	52	&	45	\\
	&	Winonaite	&	2	&	2	&	6	&	6	\\
Carbonaceous Chondrite (CC)	&		&	336	&	263	&	1146	&	921	\\
	&	CI	&	11	&	9	&	58	&	50	\\
	&	CK	&	20	&	17	&	46	&	35	\\
	&	CM	&	150	&	102	&	576	&	425	\\
	&	CO	&	34	&	31	&	99	&	89	\\
	&	CR	&	20	&	18	&	56	&	50	\\
	&	CV	&	55	&	44	&	171	&	142	\\
	&	Other CC	&	38	&	34	&	97	&	87	\\
	&	TL	&	9	&	9	&	48	&	48	\\
Ordinary Chondrite (OC)	&		&	335	&	261	&	920	&	688	\\
	&	H	&	100	&	76	&	263	&	195	\\
	&	L	&	159	&	121	&	454	&	331	\\
	&	LL	&	67	&	56	&	188	&	150	\\
	&	Ungrouped OC	&	9	&	8	&	15	&	12	\\
Enstatite Chondrite (EC)	&		&	22	&	14	&	92	&	68	\\
	&	E	&	10	&	2	&	36	&	12	\\
	&	EH	&	7	&	7	&	35	&	35	\\
	&	EL	&	5	&	5	&	21	&	21	\\
R Chondrite (RC)	&	R	&	9	&	8	&	22	&	19	\\
Iron (I)	&		&	19	&	10	&	57	&	26	\\
	&	Coarse Octahedrite	&	1	&	1	&	5	&	5	\\
	&	Coarse Octahedrite IAB	&	1	&	0	&	2	&	0	\\
	&	Finest Octahedrite	&	1	&	0	&	4	&	0	\\
	&	IA	&	5	&	4	&	11	&	8	\\
	&	Medium Octahedrite IIIAB	&	1	&	0	&	2	&	0	\\
	&	Other I	&	10	&	5	&	33	&	13	\\
Stony Iron (SI)	&		&	11	&	5	&	61	&	11	\\
	&	Mesosiderite	&	6	&	4	&	17	&	8	\\
	&	Pallasite	&	4	&	1	&	38	&	3	\\
	&	Unique SI	&	1	&	0	&	6	&	0	\\
Total	&		&	1079	&	858	&	3371	&	2674	\\
\bottomrule
\end{tabular*}
\end{table}

To classify the meteorite data, we organize and group the classifications provided in the RELAB database (Type1 and SubType from the sample catalog file) into the types and subtypes listed in Table~\ref{Table:NType1}. If there were very few spectra of a given subtype, the classification was unclear or listed as ungrouped or unusual in RELAB, the data were grouped into the "other" category of each meteorite type.
We group all descriptions of subtype broadly.  For instance, we neglect petrographic types so H3 would be classified simply as H. All spectra in RELAB labeled as "Ureilite Anomalous Polymict" are samples of Almahata Sitta and are relabeled as such in this work. Tagish Lake is labeled as "TL" and grouped within Carbonaceous Chondrites.

\subsection{Corrections applied to asteroid spectra} \label{Sec:corrections}
Asteroids that have low albedos and are observed at high temperatures (while they are close to the Sun) often emit a significant amount of flux at near-infrared wavelengths starting around 2 $\mu$m. Because this is emitted light, rather than reflected sunlight, asteroid spectra that displayed thermal excess beyond 2 $\mu$m were modeled and the thermal component was removed before comparison with RELAB samples. We applied the near-Earth asteroid thermal model (NEATM) developed by \citet{Harris1998} using a model by \citet{Volquardsen2007} and implemented by \citet{Moskovitz2017} as per the procedure described in \citet{Binzel2019} and \citet{Marsset2022}.

Space weathering is a term describing a number of processes that affect the surfaces of airless bodies due to high energy particles and micrometeorite impacts \citep{2001JGR...10610039H,Brunetto2015}. The spectral effects of these processes depend on the composition of the body. Space weathering on olivine-rich bodies such as the S-complex and A-types appears to be relatively consistent across bodies, seen mainly as a spectral reddening and darkening effect and has been well-characterized in the laboratory \citep[e.g.,][]{1999EP&S...51.1255Y,2005Icar..174...31S} and confirmed by sample return \citep{2006Natur.443...56H,Nakamura2011}. Models have been created that can correct those space-weathering effects for better direct comparison with meteorite spectra.  For other spectral types (C-complex, X-complex, etc.) the effects of space weathering are less well-understood and the spectral changes are not consistent: for some bodies the effect is spectral reddening, for others it is blue-ing \citep[e.g.,][]{2015Icar..254..135M,2017Icar..285...43L,2018Icar..302...10L}.

For asteroids in the S-complex (S, Sa, Sq, Sr, Sv, Q) and A-types, we applied a space weathering correction before comparison with meteorites. We model each spectrum using the Shkuratov model \citep{Shkuratov1999} and space weathering model \citep{Brunetto2006} using code implemented by \citet{Vernazza2008} to determine the best Cs weathering parameter, which is a value that defines the strength of the exponential curve that models weathering. We then remove the weathering component from the spectrum according to the Brunetto model. In this work, we compare these weather-corrected asteroids and meteorites over the 0.45-1.9 $\mu$m range where the weathering correction is most accurate \citep{Vernazza2008}. 

The phase angle of the observation also affects the slope of a spectrum. However, we do not apply any correction here because it has been shown that there is no single phase correction that is applicable to all objects and that the correction may depend on the individual object \citep[e.g.,][]{2018P&SS..157...82P,Binzel2019}.

\subsection{Biases and Incompleteness of the Datasets}\label{Sec:Bias}
There are a number of biases and limits to our analysis, and we acknowledge those that we most recognize here. Among meteorites, the known bias starts with delivery to the Earth's surface - realizing some locations in the main belt are less likely to be delivered to Earth \citep{Bottke2002, Greenstreet2012, Granvik2018}, the meteorite flux is time dependent \citep{Heck2017}, and materials of different strengths are more or less likely to survive entry in the atmosphere. It is conceivable that there are biases earlier in the meteorite delivery process as well: \citet{Rivkin2014} argued that ice-rich objects like Ceres might not produce families as readily as rockier objects, and \citet{Rivkin2019} noted the lack of C-complex near-Earth asteroids compared to calculations based on main-belt populations and published delivery efficiencies.

The RELAB database itself is not fully representative of the existing meteorite collection. Some meteorite types are more comprehensively measured in RELAB than others. For example RELAB does not have any samples of the following: K (Kakangari) chondrites, Interplanatery Dust Particules (IDPs), CB and CH carbonaceous chondrites, and various iron subtypes. Even though the RELAB database misses some meteorite types, it is still the largest available database and using a single dataset was chosen for coherence.

Another challenge with large-scale comparisons of the RELAB dataset is that there is duplication of spectra in some cases and varying numbers of spectra taken under different conditions for each meteorite sample. An example of duplicate (essentially identical) spectra include Spectrum IDs CGN151 and MGN151 (Sample ID: MR-MJG-076), two spectra of LL5 Paragould. In this case, the difference is just a small shift in wavelength (about 17 angstroms),  a small correction applied to the majority of this set of meteorite spectra, a total of 151 samples, as described in Gaffey\_Spectra.txt within the RELAB catalogs. An example of similar spectra taken under different conditions includes BKR1MT187 and BKR2MT187 (Sample ID: MT-JMS-187), two spectra of particulate Nogoya CM2, obtained in two different atmospheric conditions (ambient and dry air) with the Thermo Nexus 780 using two different apertures (55 and 150).

\begin{figure}
	\centering
		\includegraphics[width=1\textwidth]{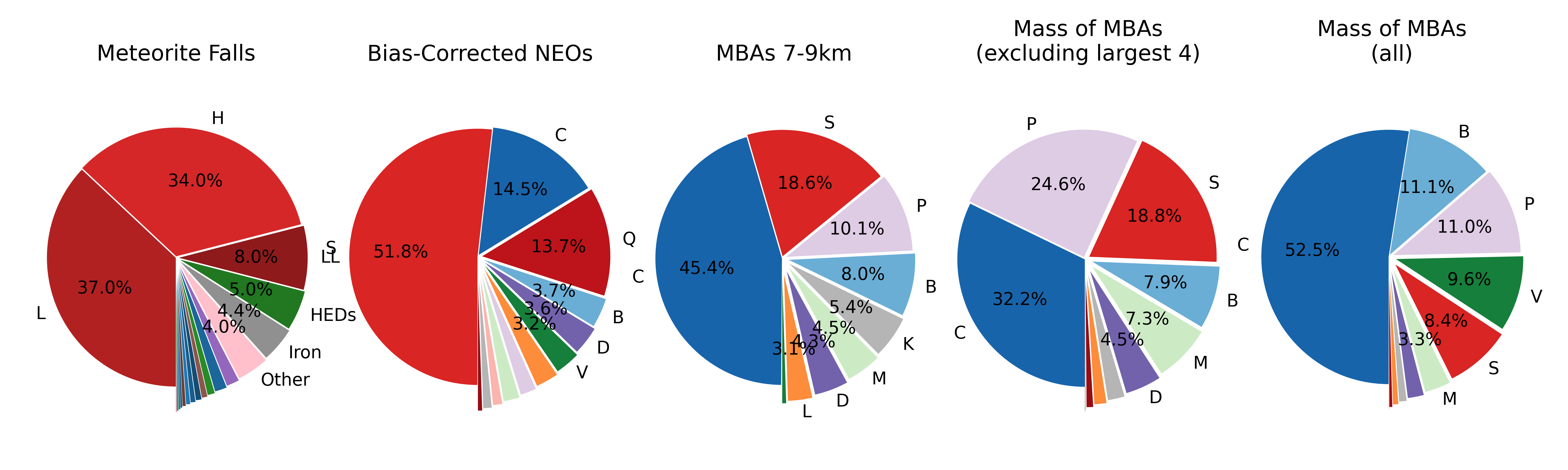}
	\caption{Fractional makeup of meteorite subtypes among meteorite falls compared with the fractional makeup of asteroid types among NEOs, small MBAs in the 7-9 km range, and the mass of asteroids excluding the largest four (Ceres, Pallas, Vesta, Hygiea), and the mass of all asteroids (including the largest four which account for 50\% of the mass). The differences between populations highlights the importance of delivery biases from the Main Belt to the NEO region to meteorites on Earth. }
	\label{FIG:piechart}
\end{figure}

There are also biases in the asteroid spectroscopic dataset. Particularly because we restrict the sample to a high signal-to-noise threshold, the dataset is dominated by large, bright main-belt asteroids and small, nearby, bright near-Earth asteroids. Fig.~\ref{FIG:piechart} shows the compositional differences between meteorites delivered to Earth, the NEO population and full Main Belt population. This emphasizes the existence of many biases in the meteorite collection and that it is not fully representative of the bodies in the asteroid belt.

\section{Asteroid-Meteorite Matching Methods}
Here we describe our method for identifying meteorites that are suitable matches to the asteroid data.

\subsection{Data normalization and smoothing} 
\begin{figure}
	\centering
		\includegraphics[width=1\textwidth]{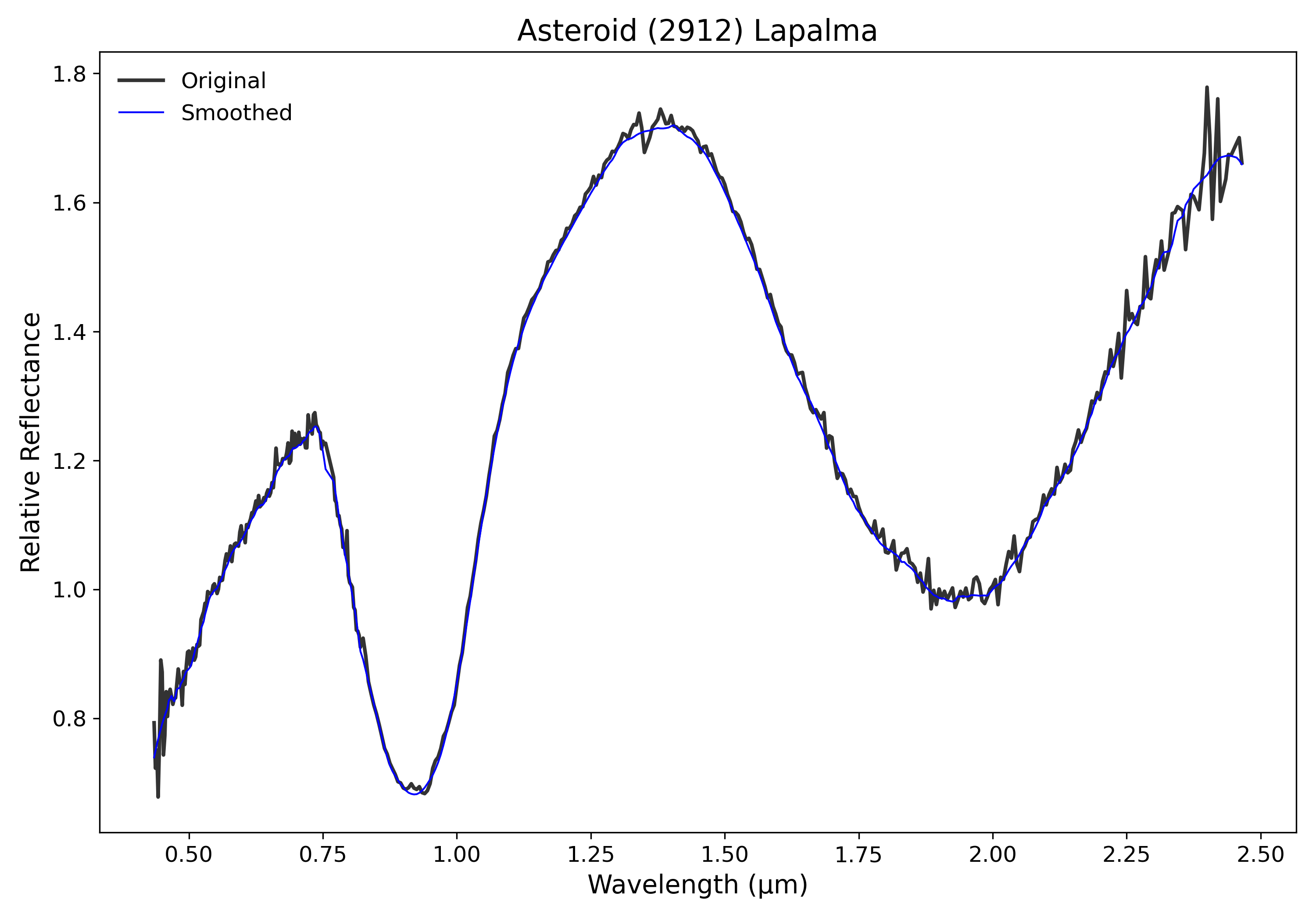}
	\caption{Typical example of original (black) and smoothed (blue) asteroid spectrum using Savitzky-Golay filtering. }
	\label{FIG:smoothing}
\end{figure}

First, each asteroid and RELAB spectrum is normalized to 1 at 0.55 $
\mu$m. We then smoothed the asteroid spectra using Savitzky-Golay filtering, fitting a cubic polynomial through the window length (polyorder=3, window\_length=21 with the python \texttt{scipy.signal.savgol\_filter}  (\url{https://docs.scipy.org/doc/scipy/reference/generated/scipy.signal.savgol_filter.html}) function. All other optional parameters were left at their default).After testing a number of different lengths, we chose a window length of 21. We found that a smaller window (fewer points) sometimes under-smoothed and created artifacts while a larger window (more points) sometimes over-smoothed and removed true features. Additionally, an odd-number is required for the window length. See Fig.~\ref{FIG:smoothing} for an example. The goal of this smoothing  was to increase the SNR of the spectra while retaining all important spectral features. We tested polynomial fits to smooth the spectra but we rejected this method because these fits did not preserve the spectral features.

The RELAB meteorite spectra generally have much higher resolution (more data points) and a larger wavelength range than the asteroid spectra. For each asteroid, each meteorite spectrum was interpolated to the same wavelength range as the asteroid spectrum.  Therefore, the result of interpolating a meteorite spectrum is essentially a lower resolution spectrum to match the resolution of the asteroid spectrum. For interpolation, we used the \texttt{scipy.interpolate.interp1d} (\url{https://docs.scipy.org/doc/scipy/reference/generated/scipy.interpolate.interp1d.html}) function. We chose a cubic interpolation, and left all other optional parameters at their default values.

\subsection{Best-fit method} 
We use the $\chi^2$ test statistic to measure goodness-of-fit of the RELAB sample data to the observed asteroid data. This method has been used in numerous studies for asteroid-meteorite comparisons and for asteroid classification \citep[e.g.][]{2010JGRE..115.6005C,2011Icar..216..462C,2012Icar..218..196D,Popescu2012}. We implement the \texttt{scipy.stats.chisquare} (\url{https://docs.scipy.org/doc/scipy/reference/generated/scipy.stats.chisquare.html}) function with all default values to perform this comparison.  

We restrict the dataset of asteroid-meteorite pairs to the 20 RELAB samples with the lowest $\chi^2$ values for each asteroid. We define a distinct sample as the ten first digits of the RELAB Sample ID. We choose 20 samples because the value is large enough to be representative yet small enough to be manageable and the results inspectable.  We choose to restrict the dataset by sample rather than by distinct spectra because different samples have varying number of spectra. Restricting the dataset by spectra would cause some samples to be missed because duplicates of the same sample would take up the lowest chi-square positions. The first 20 meteorite samples matching to the 500 asteroids results in 12,657 asteroid-meteorite spectral pairs with between 20 and 41 pairs per asteroid.
 
\begin{landscape}

\begin{figure*}
	\centering
		\includegraphics[width=1.3\textwidth]{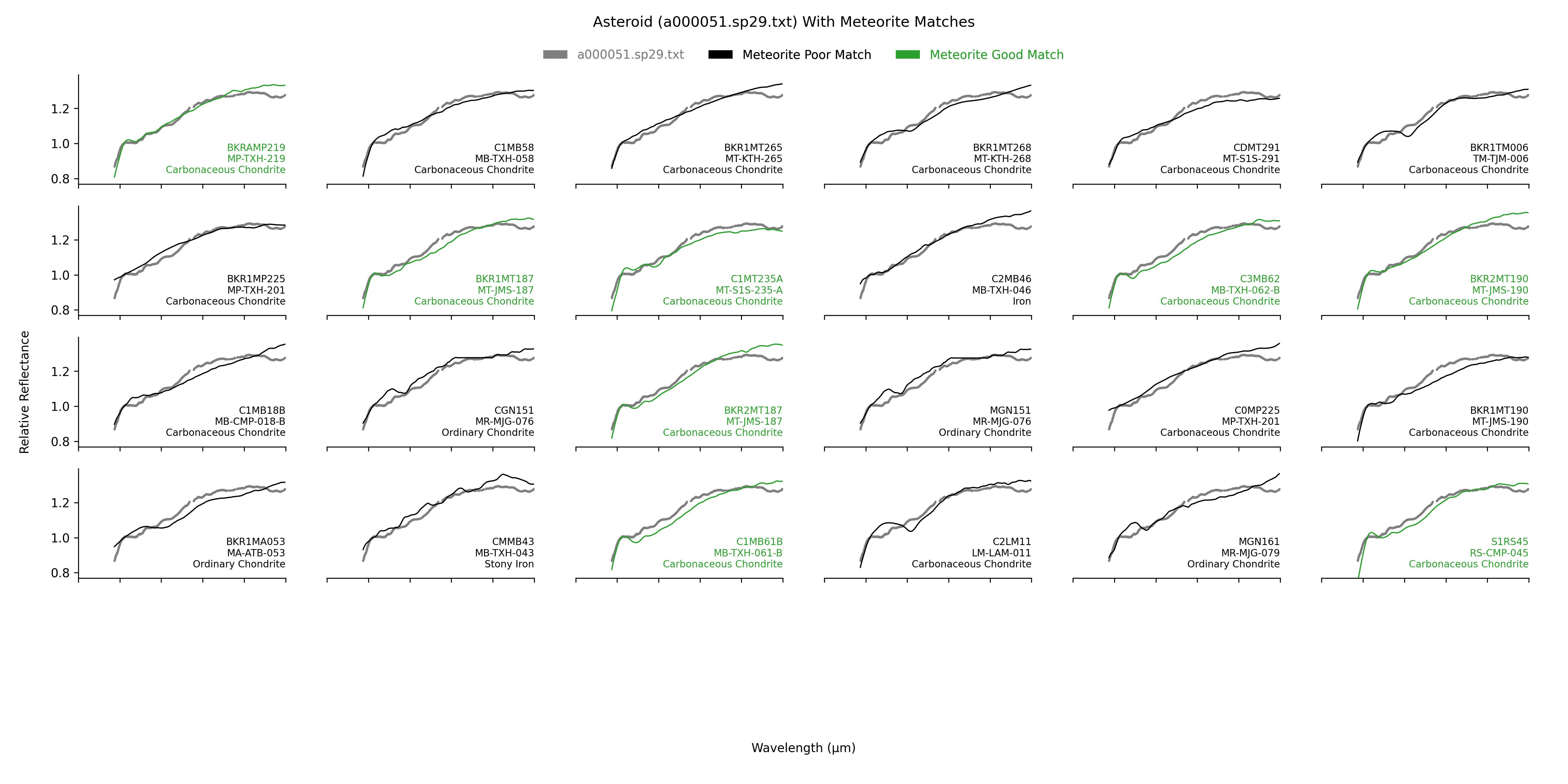}
	\caption{Example of the lowest $\chi^2$ meteorite fits to Cgh-type asteroid (51) Nemausa. In this example, good matches to a Cgh-type must have a 0.7-micron hydration feature. Meteorites plotted in green were approved through visual inspection. Those plotted in black were rejected due to mismatching absorption features. On the bottom right of each plot the meteorite Spectrum ID, Sample ID, and Type is shown. }
	\label{FIG:examplematch}
\end{figure*}
\end{landscape}

 We then visually examined the asteroid-meteorite pairs of the 20 RELAB samples with the lowest $\chi^2$ values for each asteroid.  We exclude any meteorite spectra that are not good fits through qualitative inspection that focuses on the presence of absorption features in both the asteroid and meteorite spectra and the shape and center of those absorption features and spectral shape of the continuum and any curvature.  The examination process was performed entirely by the first author and multiple steps were taken. First, we ran a code that displayed each asteroid-meteorite pair matched one at a time, batching by asteroid class and by asteroid, so all the same spectral class was reviewed together. The code allowed to click “Yes” or “No” to keep or reject the pair and would record the result so the process could be done efficiently. Second, for a random sample of asteroids in each class, usually the first few and a few others, we then reviewed the results for each asteroid with all meteorites in a single plot with visual inspection results of good matches in green and poor ones in black, as shown in Fig.~\ref{FIG:examplematch}. A small number of pairs were changed from “Yes” to “No” or the reverse, but generally, the inspection at this point was to validate that the matches looked correct. Finally, we performed additional close inspections of any matches that had not been previously found (many of which were rejected, see Sec.~\ref{Sec:rejected}), were more rare, and of asteroid classes with weaker features that warranted more careful attention.

 All meteorite-asteroid pairs kept after this inspection are considered a "match" in this work. We show an example of $\chi^2$ and visual inspection results in Fig.~\ref{FIG:examplematch}. We provide the full list of $\chi^2$ matches in the Supplementary Data. The table includes the asteroid and asteroid class, the meteorite Spectrum ID and Sample ID as well as other relevant information from RELAB, the $\chi^2$ values and whether or not the match was considered a good fit.

\subsection{A note on the limits of the methods}
Mathematically, the lower the $\chi^2$ value, the better the match. However, the method here equally weights all parts of spectrum even though there are specific spectral features that are more important, but are not more heavily weighted by the $\chi^2$ calculation. For example, an asteroid-meteorite pair that matches a specific weak spectral absorption may be the best composition match even if another meteorite has a lower $\chi^2$ value but does not exhibit the absorption feature. Thus the presence or absence of distinct spectral features is an essential condition in registering spectral matches. While qualitative visual inspection has subjectivity, the alternative is designing a set of quantitative measurements including absorption band parameters, spectral slope, and spectral shape. These measurements would need to be custom defined for every unique set of asteroid features and is beyond the scope of this work.

There are other methods that could be used to compare asteroids and meteorites. Principal Component Analysis (PCA), for example, is the most common method used for classifying asteroid spectra and could be used to compare asteroids to meteorites. Many meteorite spectra and their types scatter differently in PCA space compared to asteroids, obfuscating links \citep{Burbine2019}. PCA also creates an additional layer of challenge, because this method maximizes the variance to the fewest dimensions. Therefore, 1) the more subtle spectral features such as weak absorptions, or more subtle changes in spectral shape are not strongly distinguished among the first principal components; and 2) those principal components cannot always be linked back to specific spectral features. The first few principal components of \citet{DeMeo2009taxo} represent the 1- and 2- $\mu$m band, but for this work we chose a method that should also be effective for more minor spectral features.

\subsection{Note on confidence of a match}
We intentionally lean toward broader rather than stricter matching because spectral shape and absorption feature variation depends on sample measurement conditions. It is for this reason we allow multiple matches to each asteroid instead of choosing a unique best match under the assumption that the best match is not necessarily the exclusive compatible composition match.  Additionally, we choose not to use the $\chi^2$ value as an indication of relative confidence. Instead, we choose to equally weight the confidence of all of our matches.

The challenge with any mathematical calculation is that a given composition can be represented by a range of spectra rather than a specific spectrum, thus a correct mathematical score requires defining a phase space of spectra. For example, the slope and band depth of a meteorite measurement depends on the sample preparation - the grain size and the composition of the specific sub-piece that was pulverized. Therefore, all non-compositional effects \citep[grain size, phase angle, airmass, temperature, packing/porosity, etc.][]{Sanchez2012,Reddy2015, Marsset2020,KramerRuggiu2021} would need to be corrected before the matching process. This will require a much more comprehensive data set: meteorite measurements under a range of conditions for the same material and asteroid measurements under a range of observing conditions.

A meteorite-asteroid match in this work indicates spectral similarity over visible and near-infrared wavelengths, however, we hold as a truism that a spectral match does not guarantee a compositional match. Additional measurements of other parameters such as albedo, density, and spectral measurements in additional wavelengths are required to fully validate compatibility. We discuss this further in Sec.~\ref{Sec:Discussion}.

\subsection{Note on slope corrections}~\label{sec:slopeversion}
Differences in spectral slope can significantly affect the $\chi^2$ value, causing meteorites and asteroids with similar absorption features or spectral shape to be missed. For this reason, we tested a version of the analysis where our spectral matching method followed the removal of the linear spectral slope calculated over the full wavelength range. We note that for some asteroid types the resulting meteorite type and sub-type matches were consistent with matches without slope removal. However, for other types the matches were significantly worse. For example, for K- and L-types the meteorites from the $\chi^2$ calculation were poor fits, and for Xe-types, slope removal did not help identify meteorite matches that had the 0.9$\mu$m feature. Because the results from this version of the $\chi^2$ analysis did not improve the ability to find compatible meteorite and asteroid links, we do not address these results further. For the S-complex, we do apply a space-weathering correction that removes a non-linear slope as described in Sec.~\ref{Sec:corrections}.

\subsection{Identification and removal of unrepresentative samples} \label{Sec:rejected}
Our initial results from $\chi^2$ matching and visual inspection of spectra found a number of unexpected matches among asteroid types and meteorite subtypes. For those unexpected matches, we reviewed the measurement conditions from the RELAB sample catalog and from the notes at the end of each individual .asc data file. In several cases, the meteorite was reclassified after further inspection of the data (see Table~\ref{Table:reject}).
In many cases, the measurement and sample conditions were not representative of the full meteorite or the expected surface conditions of the asteroid and we discarded the results. We list and describe some examples in the Table~\ref{Table:reject}. For Almahata Sitta, a meteorite with fragments having a wide variety of spectral characteristics, we keep all good spectral matches. In a few cases, there are unexpected matches for which the spectrum and sample conditions are representative of the meteorite and we address those throughout Sec.~\ref{Sec:Results}. If additional information about a meteorite or asteroid type, such as albedo, density, other wavelengths, excludes a link, we discuss it in Sec.~\ref{Sec:Discussion}.

\begin{landscape}
\begin{table*}[width=1\linewidth,pos=h]
\caption{Examples of Rejected or Reclassified RELAB data}
\label{Table:reject}
\begin{tabular}{p{0.1\textwidth}p{0.1\textwidth}p{0.1\textwidth}p{0.1\textwidth}p{0.1\textwidth}p{0.5\textwidth}}
\toprule
Meteorite	&	Asteroid Type	&	Example	&	Example	&	Action	&	Reason	\\
	&	Matched	&	Spectrum ID	&	Sample ID	&		&		\\
\midrule											
Aubrite	&	B-type	&	BKR1LM031F	&	LM-LAM-031-F	&	Reject	&	The measurement was of a specific dark section of the sample	\\
Heated CC/EC	&	C-complex	&		&		&	Reject	&	We remove laboratory heated or irradiated CCs and ECs.	\\
CV Allende	&	A-type	&	CIMB63	&	MB-TXH-063-HI	&	Reject	&	Heated in a glass tube with 10-5 atm H2 for one week at 1200C.	\\
Pallasite	&	X- and D-types	&		&		&	Reject	&	Slabs described in the RELAB .asc file as iron or metal, and "polished with sandpaper", one of which was "polished like a mirror". These pallasite spectra do not display the characteristic deep and broad 1-$\mu$m features of olivine clearly present in pallasites either because the specific measurement was of only iron or because the mirror polish caused those particular sample measurements to be dominated by the flux from iron.	\\
CC	&		&	BKR1MT266	&	MT-KTH-266	&	Reclassify	&	Labeled as a carbonaceous chondrite in RELAB, but we label it as 'Ordinary Chondrite', subtype='Ungrouped OC' based on communication with the Principal Investigator of the measurement (Kieren Howard, personal communication)	\\
CC	&	K- and L-types	&	BKR1PH052	&	PH-D2M-052	&	Reclassify	&	Labeled in RELAB as CM, but meteorite QUE 99038 has CV-like absorption feature, and is recommended as C2-ungrouped in the Meteoritical Bulletin database, we reclassify it as "Other CC".	\\
Iron	&		&	CSC099	&	SC-EAC-099	&	Reclassify	&	Labeled in RELAB as an iron, however it is described as "San Carlos Olivine" in the information at the bottom of the .asc file. We relabel it as Mineral Olivine and accordingly remove it from our meteorite sample.	\\
\bottomrule
\end{tabular}
\end{table*}
\end{landscape}

\section{Results} \label{Sec:Results}
\subsection{Overall Results}
We compare asteroids to the full RELAB meteorite dataset, but for much of our results and discussion, we focus on the comparisons with particulate samples of meteorites only. The texture of asteroid surfaces varies greatly across different bodies, compositions, and size ranges. The surfaces of larger bodies (greater than around 1-10 km) are generally covered with regolith (loose rock, pebbles, and dust) of varying grain sizes \citep[e.g., ][]{2000Icar..148...12P}. The surfaces of smaller bodies (a kilometer or less) are often dominated by boulders, as seen from images of asteroids (25143) Itokawa \citep{2006Sci...312.1330F}, (162173) Ryugu \citep{2019Sci...364..268W}, and (101955) Bennu \citep{2019Natur.568...55L}.  Meteorite samples described as ``particulate'' are made of grains of varying sizes (often also called ``powders'') as opposed to chips or slabs of solid rock. Differences in grain size and surface texture between meteorite samples and asteroid surfaces can hamper direct comparison, however, meteorite slabs tend to dampen diagnostic absorption features while particulate samples intensify those features. Additionally, Itokawa is covered with regolith on parts of its surface and with boulders on other parts. However, both regolith-rich and boulder-rich areas present very similar reflectance spectra \citep{2006Sci...312.1334A}.  It is for these reasons we review the results of particulate-only meteorite samples separate from the full set of samples.

Of our 500 asteroids, 457 matched to at least one meteorite sample and 43 had no matches among the full dataset. Among particulate samples of meteorites, 64 asteroids had no matches.

\begin{landscape}
\begin{figure*}
\centering
		\includegraphics[width=1.3\textwidth]{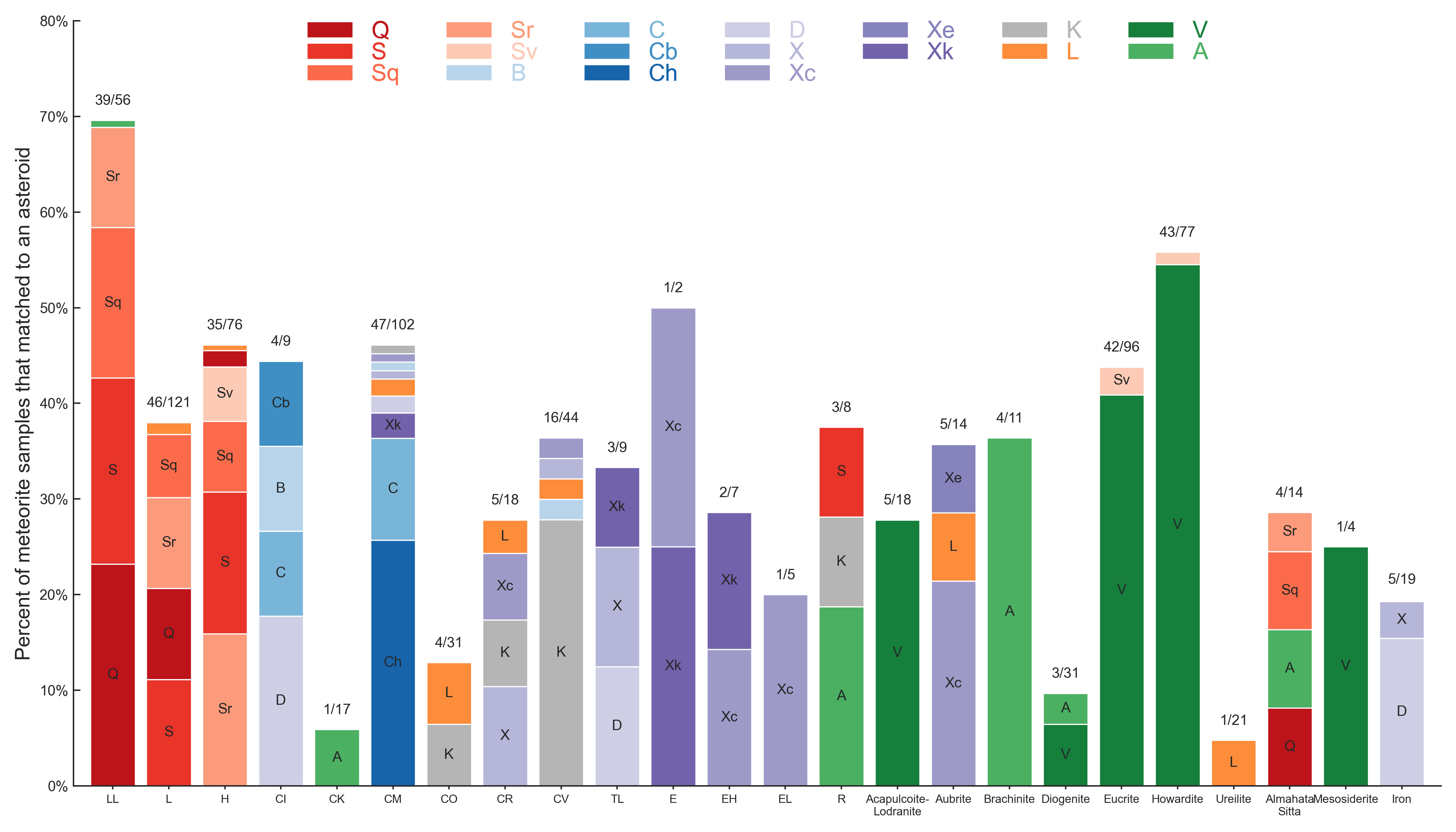}
	\caption{Comparison of asteroid types to meteorite subtypes for particulate samples only. Each bar represents the percent of RELAB meteorite samples of each subtype that matched to an asteroid spectrum, with each bar having a maximum of 100\%. Each bar displays the fraction of samples of each asteroid type that matched to the meteorite subtype. For example, 39 out of 56 LL-chondrite meteorite samples successfully matched to an asteroid. Of those samples, 31 matched to Q, 26 to S,  21 to Sq, 14 to Sr, and 1 to A. Each letter is plotted as the relative fraction of the total asteroid class to meteorite subtype matches (for LL, Q is 31/97). Matches are non-unique, there can be multiple meteorite sample matches to multiple asteroid spectra, so in some cases, for example, a meteorite sample matched to both a Q and an Sq asteroid. It is for this reason that there were 97 total asteroid class to meteorite subtype pair matches from 39 total matched samples. So 31/39 matched to Q, 21/39 matched to S and so forth.
	}
	\label{FIG:metbarchart}
\end{figure*}
\end{landscape}

Fig.~\ref{FIG:metbarchart} shows the overall results of asteroid classes that match to each meteorite subtype. This figure shows particulate samples only, a version including slabs and chips is available in the Supplementary Data.   Plotted is the percent of meteorite samples of each subtype that matched to an asteroid (of any asteroid type). Many meteorite subtypes have very strong associations with asteroid types (ordinary chondrites with the S-complex and HEDs with V-types, for example). Some associations appear more varied, for example R-chondrites with the K-, S-, L-, and A-type asteroids. However, each of those asteroid classes has a distinct 1-$\mu$m absorption feature and each class encompasses a range (varying slopes, band depths, centers, widths etc.), so the distinguishing spectral characteristics of the specific K-, S-, L- and A-types that matched to R-chondrites are smaller than their different asteroid classifications imply. Similarly for Bracchinites, most matches were to A-types, but the single matching S-type asteroid is olivine-rich and close to the A-type boarder. For Tagish Lake (TL), most matches were to D, T, and X types and the primary distinguishing characteristic among them is spectral slope.

\begin{landscape}
\begin{figure*}
	\centering
	\includegraphics[width=1.5\textwidth]{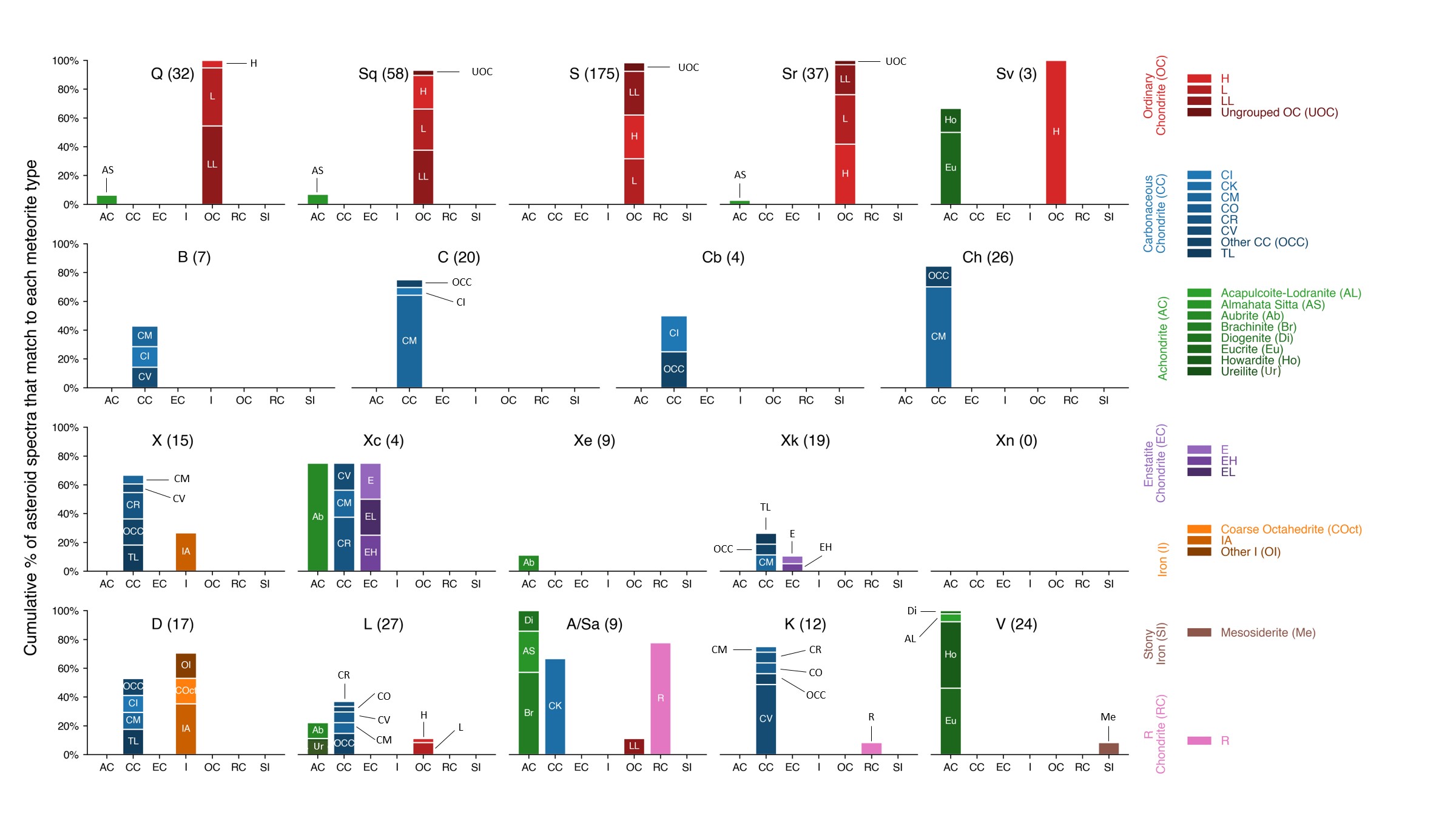}
	\caption{Percent of asteroid type matched to a meteorite type and relative contribution of meteorite subtypes for particulate samples only.  To show the strength of each connection, we plot the percent of each asteroid type that matched to a meteorite type as well as the relative contribution of each meteorite subtype.  The plot can be read according the following example: All 24 V-type asteroids matched to at least one achondrite meteorite sample so the bar height equals 100\%. Of all achondrite meteorite samples that matched V-types the subtype breakdown was: Eucrite: 42, Howardite: 42, HED Diogenite: 2, Acapulcoite-Lodranite: 5 for a total of 91 samples. Thus Eucrites and Howardites each make up 46\% (42/91) of that bar height and so forth. 	Matches are non-unique, there can be multiple meteorite sample matches to multiple asteroid spectra, however we only count distinct meteorite samples once. Each individual bar can reach a maximum of 100\%, meaning every asteroid of that asteroid type matched to a meteorite of the given meteorite type.
	}
	\label{FIG:astbarchart}
\end{figure*}
\end{landscape}

Fig.~\ref{FIG:astbarchart} shows the overall results of the matches by asteroid class for particulate samples. This figure provides the same information as Fig.~\ref{FIG:metbarchart}, but organized by asteroid class rather than meteorite class. Results including slabs and chips are provided in the Supplementary Data. To show the strength of each connection, we plot the percent of each asteroid type that matched to a meteorite type as well as the relative contribution of each meteorite subtype.

\begin{landscape}

\begin{figure*}
\centering
		\includegraphics[width=1.3\textwidth]{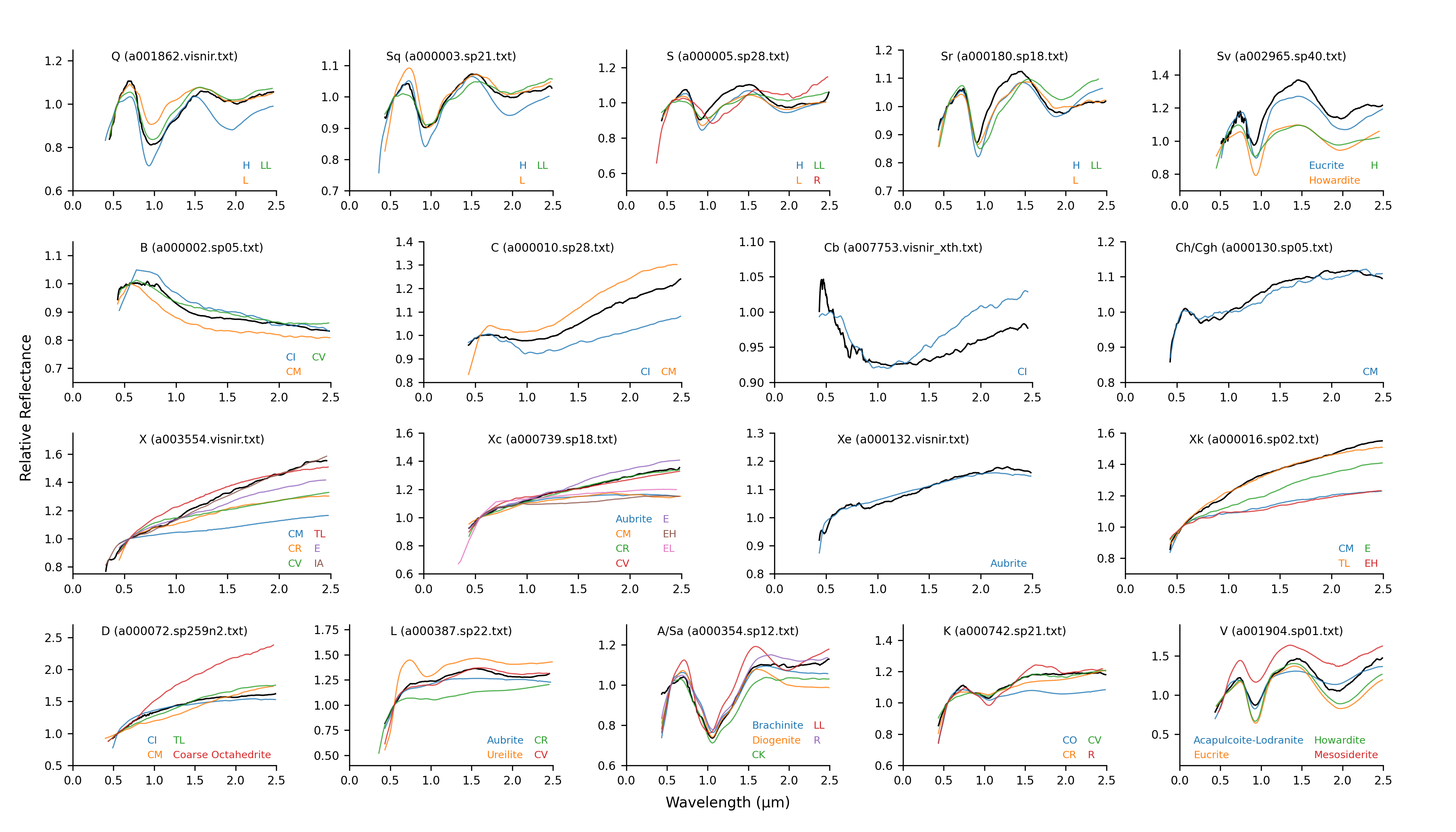}
	\caption{Asteroid and meteorite spectra that are representative of matches between classes. Plotted are a representative spectrum of each asteroid type and of each meteorite subtype match. The displayed meteorites did not necessarily match to that specific asteroid. The spectra are normalized to unity at 0.55$\mu$m. The S-complex and A-type asteroid spectra are plotted after the removal of their slope (the weathering correction). 
	}
	\label{FIG:grid_of_spectra.png}
\end{figure*}
\end{landscape}

Fig.~\ref{FIG:grid_of_spectra.png} displays representative spectra of each asteroid type and meteorite subtypes that matched to each class.

\subsection{S-complex (Q, Sq, S, Sr, Sv)}
The link between ordinary chondrites and S-complex asteroids is well established through previous meteorite and asteroid spectral studies, laboratory weathering experiments, and ground-truth of sample return from asteroid (25143) Itokawa by the JAXA Hayabusa spacecraft \citep[e.g.,][]{Binzel2001,2006Natur.443...56H,Brunetto2006,Nakamura2011}. Our results reinforce this strong S-complex - ordinary chondrite connection. Nearly all S-complex asteroids matched to at least one OC sample. Further, we can see the correlation between Q, Sq, S, Sr, and Sv-type asteroid classes with LL, L, and H meteorite classes where Q and Sq predominantly match LL and L-types OCs, S-type is intermediate with matches to all 3 subtypes, and Sr and Sv predominantly match H and L-type OCs.

As noted in previous work, ordinary chondrites are not the unique meteorite analog to the S-complex \citep{Gaffey1993, Burbine2014, Vernazza2015a}. Two out of three of Sv-type asteroids matched to samples of Howardite and Eucrite HEDs (7 samples). The Sv-type, as it is named, exhibits a narrow 1-$\mu$m and broad 2 $\mu$m feature similar to, but more shallow than, V-type asteroids. There are only three Sv-types in our sample of 305 S-complex asteroids. 

There was one match to an R chondrite (Sample ID MT-TXH-014 Spectrum ID C1MT14) and to an Acapulcoite-Lodranite (Sample ID MB-TXH-042 Spectrum ID CAMB42). Many Almahata Sitta meteorite samples were reasonable matches to S-complex asteroids. 

\subsection{C-Complex (B, C, Cb, Ch)}
Our results confirm spectral similarity between carbonaceous chondrite meteorites and the C-complex asteroids \citep[e.g.,][]{Hiroi1993, Hiroi1996}. CI samples match to C, Cb, and B-types. CV samples match to B-types.  Additionally, CM samples match to all C-complex classes.

The link between ``hydrated'' Ch (and Cgh, combined here) asteroids and CM meteorites is particularly strong because of a distinct 0.7-$\mu$m feature evident in both. The link between Ch asteroids and CM meteorites has been thoroughly explored in the literature \citep{1989Sci...246..790V,Hiroi1993, Hiroi1996, Burbine1998, Carvano2003, Takir2013, 2014Icar..233..163F, Rivkin2015}. There are cases, such as meteorite Cold Bokkeveld, where different spectra of a single meteorite match to the Ch and Xk classes. When the 0.7-$\mu$m feature is strong, it is comprised of multiple minima spanning 0.6-0.9 microns. In cases where the meteorite matched to the Xk class, the hydration feature was stronger at the minimum near 0.9 $\mu$m.

Not every CM chondrite, however, displays a 0.7-$\mu$m band due to the mineralogy and iron oxidization state of hydrous minerals.  Grain size also affects the strength and even the presence of the band because light needs to go through many phyllosillicates (usually small grain sizes) for the band to appear \citep{1973JGR....78.8507J}. In a study of thermally metamorphosed carbonaceous chondrites \citet{2012Icar..220..586C} find that as a CM meteorite is heated the 0.7-$\mu$m band weakens to depths of 2\% or less or becomes absent and that a weak silicate feature, likely olivine, begins to appear.

In most cases in this work, the presence or absence of the feature is dependant on the meteorite, although examples often have only one matched spectrum per meteorite. For example, CM meteorites Moapa Valley, MAC 88100, and Dhofar 225 match to C-types whereas ALHA 81002 matched only to Ch-types. The CM meteorites without the 0.7-$\mu$m band include both slabs and particulates. We find that most of the measurements of particulate samples of CM chondrites that don't display a strong 0.7-$\mu$m band and match to classes other than Ch are heated and have undergone alteration consistent with the results of \citet{2012Icar..220..586C} or are compositionally unique \citep[e.g.,][]{2002GeCoA..66..163K,2010M&PS...45.1108I,2013M&PS...48..879H,2014GeCoA.126..284T,2021GeCoA.298..167K}.

\citet{Vernazza2015} show that other than the CM-Ch link, additional data do not support C-complex asteroid surfaces being predominantly compositionally linked to thermally metamorphosed Carbonaceous Chondrite (CC) meteorites due to the infrequency of thermally metamorphosed CC falls ($\sim$0.2\% of meteorite falls), the difference in density, and the difference in mid-infrared spectral properties.  Other than the 0.7-$\mu$m band, diagnostic absorption bands for the C-complex asteroids are found beyond 2.5 $\mu$m and are due to hydroxylated, hydrated, ammoniated, and organic or carbonate species. \citet{Rivkin2019b} argued that the largest C-complex asteroids had no meteorite analogs based on their spectra in the 3-$\mu$m region. They note that matches based on featureless spectra in the 0.45--2.5-$\mu$m region are notional and may be superseded by data from other wavelength regions. Preliminary results from Hayabusa2 find that while Ryugu has a composition similar to CI chondrites, it has a lower albedo, higher porosity and is more fragile \citep{2021NatAs.tmp..265Y}.

There were a few matches to meteorites beyond Carbonaceous Chondrites. We find a similarity to a single Almahata Sitta chip sample. We removed most ordinary chondrite spectra that match to the C-complex from the $\chi^2$ fit after determining the particular spectra were of specific dark locations of the sample (see Sec.~\ref{Sec:rejected}).  Shock darkening is a process known to decrease the strength of silicate absorption bands and lower the albedo of the material \citep{Britt1991,Reddy2014,Kohout2014,Kohout2020}, causing the spectrum to look more C-type-like than S-type-like.  Black (shock darkened) chondrites account for nearly 14 percent of ordinary chondrite falls \citep{Britt1991}. They could represent a small fraction of small, NEO C-complex bodies, but do not represent larger, main-belt C-complex since it would be unfeasible to shock the entire surface of large bodies.

\subsection{X-Complex (X, Xc, Xe, Xk, Xn)} \label{Sec:x}

The X-complex is defined in the Bus-DeMeo system based on weak spectral absorption features at visible and near-infrared wavelengths. X-complex asteroids are also distinguishable by albedo. \citet{Tholen1984} separated these asteroids into E, M, and P that have albedos ranging from greater than 0.3 for E to intermediate for M, to less than 0.075 for P. Accordingly, we find meteorites of varying reflectivity that are spectral matches to these asteroid classes. The spectral matches with low reflectivity included carbonaceous chondrites and enstatite chondrite samples. The spectral matches with moderate reflectivity were comprised of iron samples. 

X-types with moderate, flat spectral slopes and usually no curvature matched most often with carbonaceous chondrites, but also matched with iron and one ordinary chondrite sample (C1LM05, LM-LAM-005). All ordinary chondrite samples that matched to the X-complex were measurements of slabs.

The Xc-type has concave curvature and low to moderate spectral slopes. This type matched primarily to aubrites, carbonaceous chondrites and enstatite chondrites. The link between the X-complex and enstatite chondrites and aubrites has been previously established \citep[e.g.,][]{Zellner1975, Zellner1977, Vernazza2009, Vernazza2011, Shepard2015}. The aubrites matching to Xc-types all have very high reflectance at 0.55 $\mu$m, greater than 0.3. A number of ECs matched the spectra, although they have distinct features shortward of 0.5 $\mu$m, a wavelength range not covered by our asteroid spectra. 
The Xe-type is defined by an absorption band shortward of 0.55 $\mu$m \citep{Bus2002b}. This feature has been proposed to be due to oldhamite \citep{Burbine2002} or to Ti3+ crystal field bands or Fe2+ to Ti4+ charge-transfer transitions in pyroxenes \citep{Shestopalov2010}. These spectra often also have a weak absorption near 0.9 $\mu$m similar to Xk. We only include meteorite matches here for Xe that have the band shortward of 0.55 $\mu$m, although there are very few samples that display this feature. The few that do display it are carbonaceous chondrites, enstatite chondrites, and one aubrite. All but the aubrite were measurements of slabs. 

We find examples of many meteorite types and subtypes that have spectral features - a weak 0.9-$\mu$m feature, moderate slope - consistent with the Xk class, including carbonaceous chondrites, ordinary chondrites, and enstatite chondrites.  Interestingly, four ordinary chondrite samples matched to the X-complex. Most are labeled as dark, one was irradiated with a laser, all of these meteorite spectra had muted bands, much more shallow than those that match S-type spectra and were measured from slabs or chips rather than small grain particulates. The reflectance of these meteorite spectra at 0.55 $\mu$m is less than 0.11 for all of them.

\citet{Vernazza2009} linked mesosiderites to Xk-types, particularly noting Vaca Muerta. While Vaca Muerta is in the RELAB sample it did not surface as a top match for Xk-types in this study. Reasons include the paucity of mesosiderite measurements available in RELAB and spectral slope of Vaca Muerta itself. We note, however, that even in the slope-removed version of this study (Sec.~\ref{sec:slopeversion}) Vaca Muerta was not a top match according to the $\chi^2$ analysis. This implies that a careful individual study of each meteorite type could produce additional links missed in this broad analysis. The Xk-like spectra of mesosiderites are primarily taken of slabs, whereas the V-type-like mesosiderite spectra (see Sec.~\ref{Sec:vtype}) are primarily taken of particulate samples.

\subsection{D-types} \label{sec:d}
D-type asteroids are spectrally unique due to their steep spectral slope \citep{Tholen1984}. They generally have either no absorption features or weak ones near 0.6 or 0.9 $\mu$m \citep{Vilas1993}. 
They may have slight concave curvature or be relatively flat. Tagish Lake was the earliest meteorite spectral match in the literature \citep{Hiroi2001, Barucci2018}, although they have also more recently been associated with the Tarda carbonaceous chondrite \citep{Marrocchi2021} and a primitive clast from the Zag ordinary chondrite meteorite \citep{Kebukawa2019}. While the study of Tarda measures compositional similarity to the expected composition of D-types, it does not present data to prove spectral similarity. D-types have been suggested to be rich in organics \citep{Gradie1980}, though more recent work showed abundance of $^{13}$C-rich carbonates in Tagish Lake is instead consistent with a body that accreted $^{13}$C-rich CO$_2$ ice \citep{Fujiya2019}. Recent mineralogical modeling results included  low-iron olivine, magnesium saponite-dominant phyllosilicates, and opaque material such as pyrrhotite and tholin \citep{Gartrelle2021b}. Other studies have found inner solar system analogs to D-type spectra such as Martian-moons Phobos and Deimos with lunar space-weathered anorthosite \citep{Yamamoto2018}.  

We found the spectral slope of our D-type asteroids could be reproduced by CM and CI carbonaceous chondrites, Tagish Lake, irons and enstatite chondrites. Some of our D-type asteroid spectra had a weak 0.9-$\mu$m feature, but our best matched meteorite spectra did not. This weak feature for D- and Xk-types is located at the boundary of the visible and near-infrared portions of the spectrum that were taken at different times from different instruments and at edge wavelengths where the detectors are less sensitive and thus the data are less reliable. It is likely that for some spectra this feature is an artifact, and for that reason we kept meteorite matches that did not have this feature. Asteroid (269) Justitia could not be matched due to its very steep spectral slope \citep{Hasegawa2021}. There was a single spectrum (BKR1MT020) that reproduced the slope but was rejected because it was a laboratory heated sample. Asteroid (908) Buda could not be matched due to both a distinct 0.9-$\mu$m absorption and strong spectral curvature. 

In a comparison of spectra from 0.4-25$\mu$m, \citet{Vernazza2013} found that bodies similar to Tagish Lake account for a small percent of MBAs ($<$4.5\% of the total population), and noted a lack of asteroid analogs for the Tagish Lake meteorite in the mid-infrared. Additionally, \citet{2018Icar..302...10L} show that space weathering would create a blue-ing effect on the spectrum, lowering the spectral slope. This could cause the number of bodies similar to Tagish Lake in the asteroid belt to be under-counted because they would be spectrally classified as X- or P-type rather than D-type. This result is also supported by results from \citet{2022ApJ...924L...9H} who find the spectral slope of (596) Sheila increased after the surface was refreshed from a collision.

It is surprising that a larger percentage of D-type asteroids matched to iron meteorites than X-types. The main spectral difference between the two classes is slope. The spectral slopes of iron meteorites can be affected by grain size and phase angle \citep{2010M&PS...45..304C}. \citet{2010M&PS...45..304C} find that composition changes of Fe and Ni have minimal effect, while \citet{2021PSJ.....2...95C} find the abundance of troilite can affect spectral slope. However, the link between D-type asteroids and iron meteorites can be excluded in many cases by albedo \citep{Mainzer2011}, density \citep{Carry2012}, and radar albedo \citep{2008Icar..198..294B}. D-types typically have low albedos around 0.05 with the largest visible+near-infrared D-type albedo measured to be 0.11 \citep{Mainzer2011, DeMeo2014D}, whereas the iron meteorites with similar slopes have higher reflectance values at 0.55 $\mu$m. Enstatite chondrites that matched to D-types all had albedos less than 0.1. We emphasize these matches were to enstatite chondrites, but not to enstatite achondite aubrites that have high reflectivities (generally between 0.3 and 0.55).

As discussed further in Sec.~\ref{disc:wavelengths}, \citet{2012Icar..221.1162V,Vernazza2015} showed that the mid-infrared spectral properties of D-types are similar to those of cometary anhydrous dust and that their 3-micron spectral properties \citep[featureless][]{Usui2019} are consistent with that interpretation, making analogies with hydrated CCs such as C, CI and TL less likely.

\subsection{V-types} \label{Sec:vtype}

The relationship between V-type asteroids and Howardite, Eucrite, and Diogenite (HED) meteorites is one of the strongest and earliest-known asteroid-meteorite connections \citep{Mccord1970,Consolmagno1977, Binzel1993, Prettyman2012}. Nearly all 24 V-type asteroids match to both Eucrites and Howardites. Only 2 V-type asteroids have matches to Diogenite meteorites, even though HEDs make up 1.6\%, 3.4\% and 1.2\% of meteorite falls, respectively, thus diogenites are not rare among HEDs. In the RELAB dataset there are 110 eucrite samples, 83 howardites, and 35 diogenite samples, 22 of which are labeled "olivine-diogenites". 

Our result is in line with results from \citet{Burbine2009} who find 0 of 7 NEO spectra are compatible with diogenites, \citet{Burbine2001} who find 0 of 21 main-belt V-types are diogenite, and \citet{Moskovitz2010} who find 1 of 29 main-belt V-types are diogenite. The single main-belt match is (21238) Panarea, a non-Vestoid located in the mid-belt with a semi-major axis of 2.52 AU.  \citet{2016MNRAS.455.2871I} also show that Panarea is of diogenitic composition, while on the contrary \citet{2017Icar..295...61L} show that this asteroid is compatible with Howardites. \citet{2017MNRAS.464.1718M} analyzed spectra of 10 V-types outside the Vesta family, finding 2 compatible with Diogenites. More recently \citet{2019MNRAS.488.3866M} studied 6 V-type asteroids and found one asteroid (2452) in the outer belt with a diogenitic composition.

Diogenite material comes from the subsurface of Vesta or a similar differentiated body, and in order to see a "diogenite asteroid" that material has to be excavated and reassembled without contamination by the more abundant eucrite crust. Thus, at the scale of 100s of meters to >1 km, it is not surprising that a hemispherically averaged asteroid spectrum doesn't show pure a pure diogenite spectral match. At the scale of meters (meteorites) it is easier to get the pure diogenite material - so we do see diogenite meteorites. A "howardite asteroid" does contain diogenite material by definition of howardite as a diogenite/eucrite mix.  While spectra in our sample are mostly chips off of Vesta and are not strongly diogenite, diogenite material is not missing among V-type asteroids. This is also evident from the V-types in the middle and outer belts that are more diogenitic and are likely derived from fully disrupted bodies as opposed to the cratering event on Vesta.

There are 8 Acapulcoite-Lodranite samples that match to 7 V-type asteroids. They display the characteristic 1- and 2-$\mu$m pyroxene bands, although the 2-$\mu$m band center of the meteorite is centered at shorter wavelengths. We do not perform any slope corrections to the V-type asteroid spectra (only for S-complex and A-types, Sec.~\ref{Sec:corrections}). NEO (1468) Zomba in particular has a very steep slope and that is not well matched by the HED spectra. Acapulcoite-Lodranites and one Achondrite sample labeled as "unique" in RELAB and grouped in "Other AC" in this work is able to reproduce the steeper spectral slope. The 2-$\mu$m band shape and center of the other Achondrite is not a good match, however.

\begin{figure}
	\centering
		\includegraphics[width=1\textwidth]{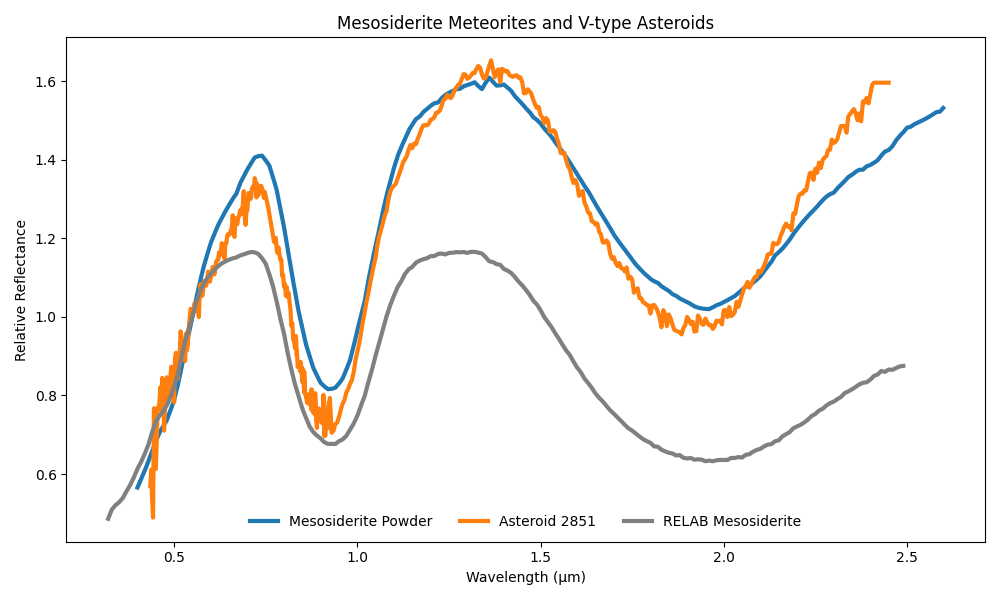}
	\caption{Comparison of mesosiderite meteorites with V-type asteroid (2851) Harbin. The spectrum labeled ``RELAB Mesosiderite'' is Spectrum ID bkr1tb148 from RELAB. The spectrum labeled ``Mesosiderite Powder'' is an average of eleven spectra taken from various mesosiderite meteorite provided by Pierre Beck. Note that Spectrum ID bkr1tb148 was not within the top 20 samples of any V-type spectra, only Spectrum ID bkr1tb151 matched. Because there were so many HED samples that match so well and few mesosiderite samples, we visually inspected all RELAB mesosiderite spectra whether or not they are included as a match in our dataset.} 
	\label{FIG:mesosiderite}
\end{figure}

We find one mesosiderite spectrum BKR1TB151 of the Pinnaroo meteorite that was similar to V-types (Fig.~\ref{FIG:mesosiderite}). While this spectrum is not as good a match to V-types as HEDs, we show an average mesosiderite spectrum that is very spectrally similar to V-types (provided by Pierre Beck, personal communication). The spectral similarity between V-types, HED meteorites, and mesosiderites have been noted in previous work \citep{2003GeCoA..67.5047W,2007LPI....38.2119B}. The HED link to Vesta and the Vesta family is firmly established, however the existence of mesosiderite meteorites suggests there could be NEOs with mesosiderite compositions.

\subsection{A and Sa-types} 

A and Sa-types are olivine-dominated spectra that display a distinct broad and deep absorption band centered near 1.05 $\mu$m \citep{Sanchez2014,DeMeo2019}. The two classes are distinguished from each other by slope, however, both have spectra indicating an olivine/(olivine+orthopyroxene) ratio of > 80\% and in many cases nearly 100\%. There is a wide range of meteorite types that display this prominent olivine band. In fact, in our comparison with the full RELAB dataset, pure olivine mineral spectra were the best matches.  Among meteorites we find many good spectral matches: pallasite Esquel, bracchinite EET99402, R-chondrite Rumuruti, CK LEW87009, diogenite MIL 03443, an LL-type OC sample, and pieces of ureilite Almahata Sitta. Pallasites, brachinites, and R-chondrites have long been known as analogs \citep{Cruikshank1984, Sunshine2007}. 

Olivine diogenite Miller Range (MIL) 03443 is a spectral match due to the olivine content, however, \citet{Beck2011} showed petrologic, geochemical, and isotopic evidence that meteorite is part of the HED clan. The match to Almahata Sitta is also due to a specific olivine-dominated, compact lithology sample of the meteorite \citep{Zolensky2010}. 

We perform our A-type matching on space-weather-corrected spectra. Uncorrected A-type spectra have steep spectral slopes due to space weathering of the large amounts of olivine \citep{1999EP&S...51.1255Y,2001M&PS...36.1587H,Sasaki2001,Brunetto2006}. Laboratory irradiation experiments of carbonaceous chondrites cause weak reddening and darkening but are unable to reproduce these steep slopes \citep{Brunetto2014}, ruling out the CK meteorite and any other carbonaceous chondrites as a true compositional match.

Among our A-types is (5261) Eureka that has a unique spectrum because the 3 minima of the 1$\mu$m band are deeper than other A-types making the whole feature more bowl-shaped than V-shaped. \citet{Rivkin2007} associated this spectrum with differentiated angrite meteorites. \citet{Polishook2017} suggest Eureka along with a group of Mars Trojans with a single dynamical source could come from Mars due to spectral similarity with Martian meteorites such as Chassigny. We did not include Martian or Lunar meteorites in this work.

\subsection{L-types}\label{sec:l}
The spectra of L-type asteroids in MITHNEOS are diverse and likely represent a range of compositions, so it is fitting that a wide range of meteorite types match to various L-type asteroids. Eight L-type asteroids do not match to any RELAB meteorites. Just over half of L-types in our sample match to carbonaceous chondrites, although the matches include various subtypes. We found some OC matches to L-type asteroids. An OC match to asteroid (402) Chloe displayed weak 1- and 2-$\mu$m features and was a laser irradiated chip. A match to asteroid (460) Scania was an OC slab that reproduced the slightly negative slope of the L-type spectrum in the near-infrared wavelength range.

Five spectra of L-type asteroids match to a single aubrite spectrum (BKR1LM031E) 
and labeled as predominantly red material of meteorite ALHA78113. The reflectance of this spectrum at 0.55  $\mu$m is 0.27, lower than typical for aubrites (generally 0.30-0.55). While the average albedo (0.149 $\pm$ 0.066) of Bus-DeMeo L-types from NEOWISE is lower than this aubrite, it is within the min-max range of 0.054 - 0.304  \citep{Mainzer2011}.

A subset of L-types, often called Barbarians after asteroid (234) Barbara, has unusual polarimetric properties \citep{Cellino2006, Gil-Hutton2008, Masiero2009,Devogele2018}. These asteroids were also suggested to be rich in Calcium Aluminum Inclusions \citep{Sunshine2008}. We do not have a large enough sample of Barbarians to comment broadly on spectral matches, however, we do find a good spectral match between (387) Aquitania and a carbonaceous chondrite CV meteorite (Spectrum ID: BKR1MP130 Sample ID: MP-TXH-130).  

\subsection{K-types} 
K-type asteroids have been linked to many carbonaceous chondrite meteorite subtypes including CV, CO, CR, and CK \citep{Bell1988,Burbine2001b,Burbine2002,Clark2009,Eschrig2021} as well the anomalous primitive achondrite Divnoe \citep{MotheDiniz2005}.  Our results agree with previous work finding matches predominantly to CV, but also finding 1-2 sample matches each to CO, CR, Ungrouped (CC) and R-chondrites. \citet{Beck2021} find that the reflectivity of CO, CR, and CK chondrites are all compatible with L- and K-type asteroids. 

\subsection{Meteorites with no Spectral Matches}
As shown in Fig.~\ref{FIG:metbarchart}, a large number of meteorite samples (611 of 1079) did not match to any asteroid. For meteorite types with lots of available data, the unmatched spectra are still spectrally similar to their associated asteroid class (ordinary chondrites to S-complex and HEDs to V-types for example). In some cases, spectra likely did not match because of the meteorite measurement conditions. For example, spectra of slabs and chips have muted bands compared to particulates whereas asteroid surfaces are generally particulate and their spectra generally match better to particulate samples. Other reasons include alteration through 1) terrestrial weathering of meteorite samples, evident in their UV, 0.5-$\mu$m, and 0.9-$\mu$m absorption bands and 2) alteration in the laboratory through heating and irradiation and 3) measurements of localized areas of a meteorite that are not representative a larger body.

\begin{figure}
	\centering
		\includegraphics[width=1\textwidth]{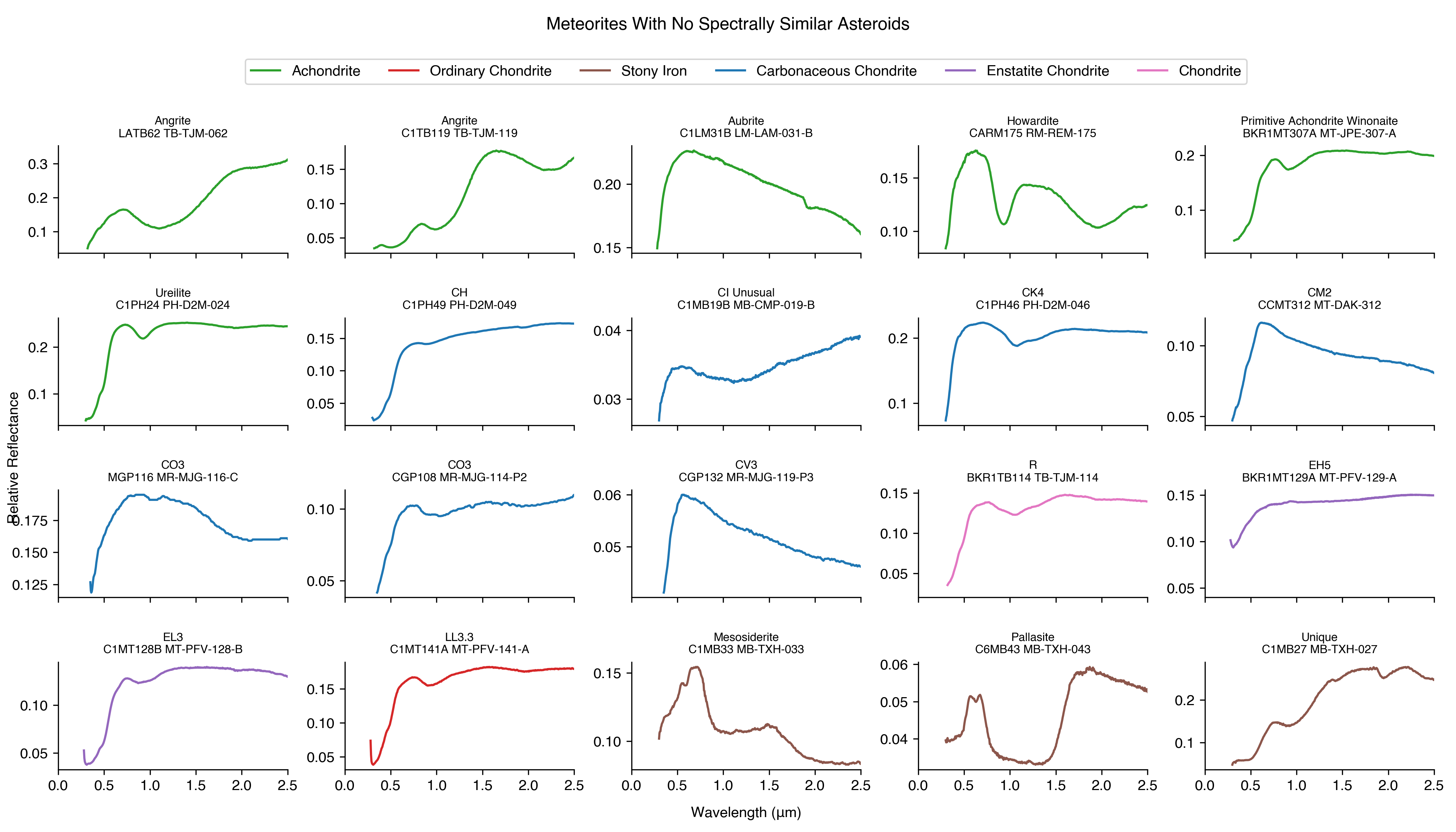}
	\caption{Plot of example meteorite spectra with spectral characteristics that did not match to any asteroids in this work. Listed above each plot is the meteorite subtype, Spectrum ID and Sample ID. Some have unusual or atypical spectral characteristics, some have flatter or slightly negative slopes not typical of asteroids, others likely did not match due to sample preparation or alteration (slab v. chip, heating or irradiation).}
	\label{Fig:metummatched}
\end{figure}

We plot some unmatched spectra in Fig.~\ref{Fig:metummatched} as examples of spectral characteristics that did not match to asteroids. Slope mismatches is certainly a contributing factor to some meteorite spectra not matching to asteroids, although removing slope before performing the analysis did not affect the overall results (Sec.~\ref{sec:slopeversion}). Some carbonaceous and enstatite chondrites had steep dropoffs at the visible wavelength end of the spectrum (also called "UV dropoff") and relatively flat spectra or negative spectral slopes not seen among asteroids. There are some unusual spectra that have absorption features not seen among our asteroid spectra, for example some angrites and a unique stony iron.  

Ureilites had very few asteroid matches, due to both the relatively steep UV dropoff and a moderately strong 0.9-$\mu$m feature. Winonaites have a spectral resemblance to the S-complex, but were not a match due to differences in the shape of the 2-$\mu$m band and to the abundance of ordinary chondrite measurements that were excellent matches to most S-complex asteroid spectra.

While some CO and CK spectra are similar to K-type asteroids with the V-shaped 1-$\mu$m feature, many spectra have a steep UV dropoff or a moderately strong 2-$\mu$m feature not seen among the existing dataset of K-type asteroids.

\subsection{Asteroids with no Spectral Matches}
Table~\ref{Table:NoMatch} lists the number of asteroids of each spectral type with no match.  Most Xe types did not have a match due to the absorption band shortward of 0.55 $\mu$m. While only 26\% of Xk types did not match among all meteorite samples, nearly 70\% did not match to particulate-only samples. This is likely because the weak 0.9-$\mu$m feature is reproduced more frequently among slabs than among particulate samples. 
L-types, and Barbarian asteroids in particular, have distinct shapes known to not be seen in the meteorite collection, but do have spectral similarities with Calcium Aluminum Inclusion (CAI) material (Sec.~\ref{sec:l}).

More than half the B-types did not have a match due to the negative slope and spectral shape. Through pulse-laser irradiation experiments of carbonaceous chondrites, \citet{2013LPI....44.1276H} and \citet{2015Icar..254..135M} found that CI/CM meteorites become bluer after irradiation. \citet{2017Icar..285...43L} and \citet{2018Icar..302...10L} also found the same trend through ion irradiation experiments of carbonaceous chondrites and comparison with featureless (B-, C-, X-, and D-type) asteroids, and that blue-ing effect could explain asteroid Bennu's blue slope. This effect could also explain why we do not find many matches to our B-type asteroids. Preliminary results from Hayabusa2 find that while Ryugu has a composition similar to CI chondrites, it has a lower albedo, higher porosity and is more fragile \citep{2021NatAs.tmp..265Y}. Additional results from samples returned from Bennu with OSIRIS-REx \citep{2017SSRv..212..925L,2019Sci...364..268W,2019Natur.568...55L} and from Ryugu with Hayabusa2 \citep{2017SSRv..208....3W,2019Sci...364..272K} will shed light on whether the C- and B-type asteroids are CI/CM meteorites or if they are compositionally consistent with Interplanetary Dust Particles (IDPs).

\begin{table}[width=.6\linewidth,cols=2,pos=h]
\caption{Number of asteroid spectra with no meteorite spectral matches}
\label{Table:NoMatch}
\begin{tabular*}{\tblwidth}{LLLLLL }
\toprule
Type	&	Asteroids	&	Unmatched	&	Unmatched	&	Unmatched	&	Unmatched	\\
	&	Total	&	All	&	All Fraction	&	Particulates	&	Partic. Frac	\\
\midrule											
A	&	9	&	0	&	0	&	0	&	0	\\
B	&	7	&	4	&	0.57	&	4	&	0.57	\\
C	&	20	&	4	&	0.2	&	5	&	0.25	\\
Cb	&	4	&	1	&	0.25	&	2	&	0.5	\\
Ch	&	26	&	3	&	0.12	&	4	&	0.15	\\
D	&	17	&	3	&	0.18	&	3	&	0.18	\\
K	&	12	&	3	&	0.25	&	3	&	0.25	\\
L	&	27	&	8	&	0.3	&	11	&	0.41	\\
Q	&	32	&	0	&	0	&	0	&	0	\\
S	&	175	&	1	&	0.01	&	2	&	0.01	\\
Sq	&	58	&	1	&	0.02	&	3	&	0.05	\\
Sr	&	37	&	0	&	0	&	0	&	0	\\
Sv	&	3	&	0	&	0	&	0	&	0	\\
V	&	24	&	0	&	0	&	0	&	0	\\
X	&	15	&	2	&	0.13	&	4	&	0.27	\\
Xc	&	4	&	0	&	0	&	0	&	0	\\
Xe	&	9	&	6	&	0.67	&	8	&	0.89	\\
Xk	&	19	&	5	&	0.26	&	13	&	0.68	\\
Xn	&	2	&	2	&	1	&	2	&	1	\\
Total	&	500	&	43	&	0.086	&	64	&	0.13	\\
\bottomrule
\end{tabular*}
\end{table}

\begin{landscape}
\begin{table}[width=1\linewidth,cols=8,pos=h]
\begin{minipage}[t]{\textwidth}
\renewcommand{\footnoterule}{}
\caption{Meteorite-Asteroid Associations}
\label{Table:vdqitable}
\begin{tabular*}{\tblwidth}{LLLLLLLLL}
\toprule
Meteorite Type	&	Fall	&	Meteorite	&	Fall	&	Example Meteorite	&	Met. Samples \footnote{Results for particulate-only samples. Example: 35 out of 76 (46.1\%) H chondrite meteorites in the RELAB dataset matched to one or more asteroids in the asteroid dataset in this work.}	&	Total & Asteroid Class \footnote{Results for particulate-only samples. Example: There were 81 asteroid-meteorite matches to 76 H-chondrite samples (meteorites can match to more than one asteroid). 28 of the 81 H-chondrite matches were to Sr-type asteroids (34.6\%). 26 of the 81 matches were to S-type asteroids (32.1\%).}		\\
	&	Frequency	&	Subtype	&	Frequency	&		&	Matched (\%)	& Matches \footnote{The total number of asteroid-meteorite matches for a given meteorite class. This number may be higher than the total number of meteorite samples for that class. For example a single LL-chondrite sample could match to an S-type asteroid and an Sq-type asteroid.}	& Matched (\%)		\\
Ord. Chondrite	&	80	&	H	&	34	&	Ochansk, Pantar, Forest City	&	46.1	&	81	&	Sr:34.6,S:32.1,Sq:16.0,Sv:12.3,Q:3.7,L:1.2		\\
	&		&	L	&	37	&	Vouille, Bald Mountain	&	38	&	92	&	S:29.3,Sr:25.0,Q:25.0,Sq:17.4,L:3.3		\\
	&		&	LL	&	8	&	Chelyabinsk, Jelica, Bandong	&	69.6	&	93	&	Q:33.3,S:28.0,Sq:22.6,Sr:15.1,A:1.1		\\
Carb. Chondrite	&	4.40	&	CI	&	0.5	&	Orgueil, Ivuna	&	44.4	&	5	&	D:40.0,Cb:20.0,B:20.0,C:20.0		\\
	&		&	CK	&	0.2	&	LEW87009	&	5.9	&	1	&	A:100.0		\\
	&		&	CM	&	1.5	&	Migei, Murchison	&	46.1	&	52	&	Ch:55.8,C:23.1,Xk:5.8,D:3.8,L:3.8,K:1.9,Xc:1.9,B:1.9,X:1.9		\\
	&		&	CO	&	0.6	&	Lance	&	12.9	&	4	&	L:50.0,K:50.0		\\
	&		&	CR	&	0.2	&	Al Rais, Renazzo	&	27.8	&	8	&	X:37.5,Xc:25.0,K:25.0,L:12.5		\\
	&		&	CV	&	0.7	&	Allende, Grosnaja, Mokoia	&	36.4	&	17	&	K:76.5,Xc:5.9,X:5.9,L:5.9,B:5.9		\\
	&		&	TL	&	0.0	&	Tagish Lake	&	33.3	&	8	&	X:37.5,D:37.5,Xk:25.0		\\
R Chondrite	&	0.10	&	R	&	0.1	&	Rumuruti	&	37.5	&	4	&	A:50.0,S:25.0,K:25.0		\\
Enst. Chondrite	&	1.60	&	EL	&	0.8	&	Pillistfer	&	20.0	&	1	&	Xc:100.0		\\
	&		&	EH	&	0.8	&	Abee	&	28.6	&	2	&	Xk:50.0,Xc:50.0		\\
Iron	&	4.60	&	Iron	&		&	Odessa, Casey County	&	26.3	&	5	&	D:80.0,X:20.0		\\
Stony Iron	&	1.10	&	Pallasites	&	0.4	&	Esquel	&	0.0	&	0	&	-		\\
	&		&	Mesosiderite	&	0.7	&	Pinnaroo	&	25.0	&	1	&	V:100.0		\\
Achondrite	&	8.00	&	Acapuloite+Lod.	&	0.2	&	Acapulco, Lodran	&	27.8	&	5	&	V:100.0		\\
	&		&	Aubrite	&	0.9	&	Mayo Belwa	&	35.7	&	5	&	Xc:60.0,Xe:20.0,L:20.0		\\
	&		&	Brachinite	&	0	&	Eagles Nest, Brachina	&	36.4	&	4	&	A:100.0		\\
	&		&	Howardite	&	1.6	&	Le Teilleul, Petersburg	&	55.8	&	43	&	V:97.7,Sv:2.3		\\
	&		&	Eucrite	&	3.4	&	Pasamonte	&	43.8	&	45	&	V:93.3,Sv:6.7		\\
	&		&	Diogenite	&	1.2	&	Johnstown	&	9.7	&	3	&	V:66.7,A:33.3		\\
	&		&	Ureilite	&	0.6	&	DaG319	&	4.8	&	1	&	L:100.0		\\
	&		&	Almahata Sitta	&		&	Almahata Sitta	&	28.6	&	7	&	Sq:28.6,A:28.6,Q:28.6,Sr:14.3		\\
\bottomrule
\end{tabular*}
\end{minipage}
\end{table}
\end{landscape}

\subsection{Robustness of Results}
Because the RELAB dataset has varying numbers of meteorite spectra per sample, we explored two ways of limiting the dataset to determine how much the results differed. Presented in this work, we limit the first 20 samples by lowest chi-square (20-sample version). This resulted in 12,657 asteroid-meteorite spectral pairs, with each asteroid having between 20 and 41 meteorite spectra matched. Using the same dataset, we restricted to the 30 spectra (instead of 20 samples) with the lowest chi-square (30-spectra version). This resulted in 15,000 asteroid-meteorite pairs (30 x 500) and the number of distinct samples varied from 14 to 30. There were only 168 asteroid-meteorite pairs in the 20-sample version that were not included in the 30-spectra version. The 20-sample version had a dataset that was 15.6\% smaller. However, of the particulate-only pairs that were considered good fits, the number of total samples that matched decreased from 327 to 317, a 3.7\% reduction. For each meteorite subtype, the difference was 0 to 2 samples, so for LL-chondrites, for example, 46 or 48 samples matched among 121 total samples (for the 20-sample and 30-spectra versions, respectively, a change from 38.0\% of samples matching an asteroid to 39.7\%). This means if we removed a sample from one asteroid of a given class when restricting to 20 samples, that sample was already a good match to another asteroid of the same class. Therefore, when counting distinct meteorite samples that matched to asteroid classes, the number of spectra or samples chosen within a reasonable range does not have a huge impact on results.

\section{Discussion} \label{Sec:Discussion}

While this work focused on visible and near-infrared spectral data compatibility between asteroids and meteorites, there are a number of other measurements that can help strengthen or weaken proposed links. Here we overview additional constraints from albedo, density, and additional wavelength ranges.

\subsection{Constraints from other measurements: Albedo}
From large infrared surveys, albedo estimates are available for greater than 100,000 asteroids \citep{Tedesco2002,Trilling2010,Mainzer2011,Masiero2011,Usui2011,Mueller2011,Hasegawa2013}, that have enabled average albedos and ranges to be defined for each asteroid spectral class.

The reflectance values of meteorites are not directly relatable to asteroid geometric albedos, partly due to viewing geometry and the opposition effect that encompasses shadow-hiding and coherent backscatter \citep[e.g.,][]{2000Icar..147...94B,Hiroi2001}. Because most meteorite measurements are taken at a phase angle of 30 degrees their reflectances are typically lower due to shadows than an asteroid taken at a lower phase angle. Most main belt asteroids are measured at low phase angles, while NEOs are measured at a wide range of phase angles that can be much greater than 30. \citet{Beck2021} determined a relation to calculate an "equivalent geometric albedo" from meteorite reflectance values that can be directly compared to asteroid albedos. This calculation corresponds to a roughly 10-20\% higher equivalent geometric albedo for reflectances greater than 0.15, but for lower reflectances around 0.05, the equivalent geometric albedo is 50\% greater.

This correction amplifies some albedo differences between meteorite subtypes. Based on an albedo comparison \citet{Beck2021} find that 1) Tagish Lake group, CI, and CM chondrites are the only carbonaceous chondrites with equivalent albedo compatible with C- and D-type asteroids, 2) the CO, CV, CR and CK chondrites have significantly higher albedos that are incompatible with C-types, 3) the CO, CR and CK chondrites have albedos compatible with L- and K-type asteroids, and 4) the albedo for ordinary chondrites trends with petrographic type and shock stage, and the albedo value of type 3 ordinary chondrites is significantly lower than that for S-types suggesting S-type surfaces may instead be more thermally processed similar to type 4 or higher \citep{Beck2021}. 

The X-complex is well-known to be spectrally degenerate and have a wide range of albedos representing a variety of compositions. We addressed those in Sec.~\ref{Sec:x}. D-type asteroids generally have low albedos (<0.1), so some of our spectral matches can be ruled out as described in Sec.~\ref{sec:d}.

\subsection{Constraints from other measurements: Density}
  The field of density measurements has dramatically improved over the last two decades, with
  numerous studies of binary asteroids 
  \citep[providing their mass, e.g.,][]{2006Natur.439..565M, 
    2014Icar..239..118B, 2021Icar..35814275D, 2020AA...641A..80Y, 
    2015aste.book..355M}
  and accurate volume determination
  \citep[through shape modeling, e.g.,][]{
    2010Icar..205..460C, 2015aste.book..183D, 2017AA...604A..64M, 2019AA...624A.121H,2020NatAs...4..569M,2021A&A...654A..56V}.
  For large asteroids with diameters greater than $\sim$100 km, macroporosity is minimal and their density can be compared to analogue meteorites and materials.
  For smaller asteroids, however, macroporosity can be as high as 50–60\% \citep{Consolmagno2008, Carry2012}. That significantly affects the measured overall density and obscures the bulk density of the rock itself. In fact, the macroporosity is generally estimated by comparing an asteroid density with the bulk density of the expected meteorite analog. Density can help place some constraints, particularly in cases of extreme high or low asteroid densities \citep{Vernazza2011,Carry2012,Weiss2012}.

  The spectral link between S-types and ordinary chondrites is well supported by the density of the largest S-type asteroids \citep{2015-AA-581-Viikinkoski, 2017AA...604A..64M, 2019AA...624A.121H},
  although the low density of (25143) Itokawa \citep[the sample of which confirmed the S-complex-OC relationship,][]{2011-Science-333-Yurimoto} 
  is explained by macroporosity \citep{2006Sci...312.1330F}.
  
  The link between hydrated C-complex (Ch and Cgh) and CM carbonaceous chondrites is also
  well supported by the density of the largest Ch asteroids with satellites
  \citep{2009Icar..203...88D, 2016ApJ...820L..35Y, 2019AA...623A.132C}.

  The asteroids belonging to the X-complex display the largest range of density
  and albedo. Several associations seem well-established.
  The high albedo X-types \citep[E-types in][taxonomy]{Tholen1989} 
  have been previously linked with enstatites, 
  which is consistent with their density
  \citep[e.g., Lutetia,][]{2011-Science-334-Sierks, 2011Sci...334..491P}.
  The low albedo X-types (also called P-types in the Tholen taxonomic system)
  are the asteroids displaying the lowest
  density \citep{2006Natur.439..565M, 2018Icar..309..134P, 2021AA...650A.129C}.
  This is consistent with a composition and structure similar to comets, and hence
  with a composition consistent with IDPs \citep{Vernazza2015}.

  The case of X-types with moderate albedo (M-types) is less obvious.
  Based on both their optical and radar albedos \citep{2010Icar..210..674O, 2014Icar..238...37N}, many
  M-types are thought to have a significant iron content 
  \citep[e.g., Psyche,][]{2013-Icarus-226-Matter, 2020JGRE..12506296E}.
  The most massive X-type asteroids (above 3--3.5 g/cc) are indeed all M-types based on albedo
  \citep{2011Icar..211.1022D, 2018A&A...619L...3V, 2021A&A...653A..57M}.
  Yet, the highest asteroid densities are below $\sim$4\,g/cc, much lower
  than that of iron meteorites that generally have densities greater than $\sim$7\,g/cc \citep{1998-MPS-33-Consolmagno}.

\subsection{Constraints from other measurements: Other wavelengths}\label{disc:wavelengths}

Comparing asteroid and meteorite spectral measurements over other wavelength ranges can provide additional information and constraints. The most studied spectral region beyond the 0.45-2.5-$\mu$m range are the 2.5-5$\mu$m where a 3$\mu$m absorption feature indicative of hydration may be present \citep{Takir2013,Rivkin2015b}
and the 8-13\,$\mu$m with silicate features
\citep{2005-Icarus-173-Lim, 2012A&A...537A..73L,Vernazza2015, 2016Icar..269...62L,2016A&A...586A..15M, Vernazza2017}. 
While no formal taxonomy has been established beyond 2.5 $\mu$m, asteroid measurements in the 3-$\mu$m region suggest a diversity of spectral shapes of low-albedo objects reminiscent of what is seen in higher-albedo objects in the 0.7--2.5-$\mu$m region.

\citet{Takir2019} and \citet{Potin2020} performed measurements of carbonaceous chondrite meteorites under vacuum and high temperature (thermally-desiccated conditions) that enabled more direct comparison with features seen in asteroid spectra. These works find that the characteristics of the 3-$\mu$m absorption feature indicate varying composition and aqueous alteration.

\citet{Usui2019} measured 66 asteroids across many spectral types in the 2.5-5-$\mu$m range.  They found that most C-complex asteroids and some of the X-complex and D-type asteroids have clear absorption features related to hydrated minerals near 2.75 $\mu$m, while no S-complex asteroids in their sample displayed absorption bands at this wavelength. They detected weak or possible absorption bands and structure between 2.5 and 3.5 $\mu$m for A-, L-, and K-type asteroids.  \citet{2011Icar..213..265G} and \citet{2014LPI....45.1092G} ) found a hydration band on (951) Gaspra through data from the Near Infrared Mapping Spectrometer (NIMS) in NASA's Galileo mission.  Therefore, hydrated material on some S-type asteroid surfaces does exist.

\citet{Vernazza2015} compiled mid-infrared spectra of asteroids in the 8-13 $\mu$m range. Their work showed that while C, X, and D asteroids and carbonaceous chondrites (aside from the Ch, Cgh to CM chondrite link) appear spectrally compatible across visible and near-infared wavelengths, at mid-infrared wavelengths their spectra are different. In fact, they find that C-complex asteroid spectra are similar to Interplanetary Dust Particles (IDPs).

Measurements of ultraviolet (UV) wavelengths can also be diagnostic. \citet{Hendrix2006,Vilas2015,Hendrix2019} measured C- and S-complex asteroids in the 0.22-0.32 $\mu$m wavelength range, identifying composition and space weathering effects. Measurements in Near-UV of space telescope GALEX (180-280 nm) also indicate distinctions between the S- and C-complexes \citep{2015ApJ...809...92W}. The Eight-Color Asteroid Survey \citep[ECAS,][]{1985Icar...61..355Z} spanned 0.33-1$\mu$m. The short end of the wavelength range enabled distinguishing spectral characteristics in the Tholen asteroid taxonomy \citep{Tholen1984} not discernible in later taxonomies such as \citet{Bus2002b} and \citet{DeMeo2009taxo}. For example Tholen classes B, C, F, and G were primarily distinguished by the 0.3–0.5-$\mu$m UV dropoff region.

\subsection{Constraints from other measurements: Spacecraft}
Measurements from spacecraft and especially sample return allow a much more in-depth characterization of an asteroid than Earth-based measurements including more definitive connections to meteorites and meteorite types.  Earth-based spectra predicted that S-type asteroid Itokawa has an LL ordinary chondrite composition \citep{Binzel2001,2006Natur.443...56H}, but confidence in such a specific asteroid-meteorite connection required the LL chondrite confirmation achieved through analysis of the Hayabusa spacecraft returned sample \citep{Nakamura2011}.  The Dawn mission studied V-type asteroid (4) Vesta and C-type asteroid (1) Ceres in great depth \citep{2011SSRv..163....3R,2013M&PS...48.2076R,2015Natur.528..241D,2016Natur.536...54D}.  Vesta’s likely link to Howardite-Eucrite-Diogenite (HED) meteorites had been proposed for decades (\citeauthor{Mccord1970}, \citeyear{Mccord1970}; see summary in  \citeauthor{2013M&PS...48.2090M}, \citeyear{2013M&PS...48.2090M}), with definitive confirmation achieved by \textit{in situ} elemental analysis achieved by instruments aboard the Dawn spacecraft \citep{Prettyman2012}.      Dawn's comprehensive measurements at Ceres have not yielded a specific meteorite connection.  Ceres appears to have a carbonaceous chondritic composition that is unlike anything in our meteorite collections \citep{2015Natur.528..241D}.

In this decade, we have had or will have close measurements of C-, D-, P-, and M-type asteroids that lack strong distinguishing spectral features in the visible and near-infrared wavelengths studied in this work. The Hayabusa2 mission has returned samples from C-type asteroid (162173) Ryugu \citep{2017SSRv..208....3W,2019Sci...364..272K}. Preliminary analyses find that Ryugu has a composition similar to CI chondrites, but has a lower
albedo, higher porosity and is more fragile \citep{2021NatAs.tmp..265Y}. The OSIRIS-REx mission will return samples from B-type asteroid (101955) Bennu \citep{2017SSRv..212..925L,2019Natur.568...55L}. Early spectral data from the mission show evidence for hydrated minerals, seen as an absorption at 2.7 microns that cannot be observed from Earth suggesting an association with aqueously altered CM-type carbonaceous chondrites \citep{2019NatAs...3..332H}. The Lucy mission will fly by seven (C-, P-, and D-type) Trojan asteroids \citep{2021PSJ.....2..172O}. The Psyche mission is set to visit M-type (16) Psyche \citep{2020JGRE..12506296E}. M-type asteroids are a moderate-albedo subset of X-types, often having high radar polarization ratios suggesting a metal-rich composition.

\section{Conclusion}
In this work we identified spectral links between asteroids and meteorites. We use $\chi^2$ fitting then visual inspection of results to identify spectral matches between 500 visible plus near-infrared (0.45--2.5 $\mu$m) asteroid spectra and over 1,000 samples of RELAB meteorite spectra. We reproduce and confirm many major and previously known meteorite-asteroid connections and find new, more rare, or less established connections as summarized here.

\begin{itemize}
\itemsep=0pt
\item  We associate ordinary chondrites with S-complex asteroids and find a trend from Q -> Sq -> S -> Sr -> Sv correlates with a stronger association with LL -> L -> H, with Q-types matching predominately to L and LL ordinary chondrites and Sr and Sv matching predominantly with L and H ordinary chondrites. 

\item We find ordinary chondrite samples that match to X-complex asteroids. These meteorite spectra had very weak absorptions, unlike most OC spectra. These were measurements of slabs and many were shocked, labeled as dark or black ordinary chondrites.   While these spectral matches are interesting, they likely only represent a minority of C- or X- complex asteroids \citep{Britt1991,Reddy2014,Kohout2014,Kohout2020}.

\item We reproduce the established link between Ch and Cgh "hydrated" asteroids and CM meteorites. We find a number of CM meteorites, most of which are heated, matched to C-complex classes other than Ch and had either no hydration feature or a very weak feature. However,  data from other work show the carbonaceous chondrites cannot be the primary link to most C-complex asteroids based on albedo \citep{Beck2021}, IR-wavelength measurements and additional spectral measurements of IDPs \citep{Vernazza2015}.

\item Among the X-complex asteroids, the strongest association is between Xc and enstatite chondrites, a well-established link. We find carbonaceous chondrites and aubrites are common spectral matches as well as some irons. Because these meteorite types have a wide range of albedos and densities - and the X-complex is also known to have a wide range of albedos - spectral data alone is not sufficient to establish exact asteroid-meteorite links. Very few meteorites displayed the 0.55-$\mu$m absorption feature characteristic of Xe-type asteroids. The few that did were carbonaceous chondrites, enstatite chondrites, one aubrite and one ordinary chondrite.
\item The predominant spectral matches to D-type asteroids were carbonaceous chondrites, including Tagish Lake, and iron meteorites. The iron meteorites have reflectances and densities higher than typical for D-types, ruling them out for the most of D-types. Any association with a specific D-type asteroid would need to have supporting albedo and density measurements. We found matches to enstatite chondrites that had reflectances compatible with the range of D-type albedos.

\item K-types matched predominantly to CV meteorites with additional matches to CO, CR and one R-chondrite. L-type asteroids have a variety of spectral characteristics. We found a few spectral matches, with the strongest associations with various carbonaceous chondrites, but no strong single association. This is consistent with \citet{Beck2021} who find that the reflectivity of CO, CR, and CK chondrites are all compatible with L- and K-type asteroids. 
\item A and Sa types matched to Brachinites, Pallasites and R chondrites, as well as to CV, and an olivine-rich LL-type ordinary chondrite. These spectra all display strong olivine absorption features.
\item V-type asteroids are clear spectral matches to HEDs in this work, as expected. We find HED matches to Sv-types that have similar although weaker absorption bands as V-types. We also find Acapulcoite-Lodranite samples that are compatible matches with some V-type asteroids, though in most cases the 2-$\mu$m band center of the meteorites are at shorter wavelengths than the asteroids. We also find mesosiderites to be spectrally compatible with V-types.
\end{itemize}

Spectral similarities between asteroids and meteorites are most firmly established when diagnostic absorption features are present in the spectrum. For the asteroid classes without any distinct features or with weak features in the 0.45--2.5 $\mu$m wavelength range (C-complex, X-complex, D-type, T-type), other wavelength ranges (2.5-5 or 8-13\,$\mu$m) where clear spectral features do exist, will help provide more definitive constraints.
 
We find in many cases the asteroid type to meteorite type links are not unique. While there are examples of dominant matches between an asteroid type and meteorite type that are well established, there are less common but still spectrally compatible matches between many asteroid types and meteorite types. This result emphasizes the diversity of compositions and highlights the degeneracy of classification by spectral features alone requiring additional measurements to firmly establish asteroid-meteorite links.

\section{Acknowledgements}

We thank Taki Hiroi for helpful discussion and Pierre Beck for providing unpublished mesosiderite spectral data. We are grateful for two anonymous referees for their very thorough reviews that greatly improved the manuscript. 
This work has used spectra obtained by investigators at the NASA RELAB facility at Brown University. We would like to acknowledge the RELAB database as a vital resource to the planetary astronomy community.
  Observations reported here were obtained at the NASA Infrared Telescope
  Facility,  which is operated by the University of Hawaii under under contract 80HQTR19D0030 
  with the National Aeronautics and Space Administration.  
  The MIT component of this work is supported by NASA grant 80NSSC18K0849. 
  Any opinions, findings, and conclusions or recommendations
  expressed in this article are those of the authors and do not necessarily
  reflect the views of the National Aeronautics and Space Administration.




\bibliographystyle{cas-model2-names}

\bibliography{metref}


\end{document}